\documentclass[journal]{IEEEtran}
\usepackage{lineno} 
\usepackage{footmisc} 
\usepackage{dblfloatfix}
\usepackage{amssymb,amsmath,amsfonts,mathtools} 
\usepackage{tabto}
\DeclareMathOperator*{\argmin}{arg\,min} 
\usepackage[ruled,vlined,linesnumbered]{algorithm2e} 
\usepackage{algpseudocode} 
\usepackage{xurl} 
\usepackage{times} 
\usepackage{fancyhdr}
\usepackage{graphicx} 
\usepackage{subfig}
\graphicspath{ {figs/} } 
\usepackage{float} 
\usepackage{cite} 
\usepackage{multicol} 
\usepackage{setspace} 

\begin{document}

\bstctlcite{IEEEexample:BSTcontrol} 
\title{Constrained Nonnegative Matrix Factorization for Blind Hyperspectral Unmixing Incorporating Endmember Independence}
\author{E.M.M.B.~Ekanayake,~\IEEEmembership{Graduate~Student~Member,~IEEE,}
        H.M.H.K.~Weerasooriya,
        D.Y.L.~Ranasinghe,
        S.~Herath,
        B. Rathnayake,
        G.M.R.I.~Godaliyadda,~\IEEEmembership{Senior~Member,~IEEE,}
        M.P.B.~Ekanayake,~\IEEEmembership{Senior~Member,~IEEE,}
        and~H.M.V.R.~Herath,~\IEEEmembership{Senior~Member,~IEEE}
\thanks{E.M.M.B. Ekanayake is with the Department of Electrical and Computer Systems Engineering, Monash University, Clayton, Victoria 3800, Australia and also with the Office of Research and Innovation Services, Sri Lanka Technological Campus, CO 10500, Sri Lanka (email: mevan.ekanayake@monash.edu).}%
\thanks{H.M.H.K. Weerasooriya, D.Y.L. Ranasinghe, S. Herath, G.M.R.I. Godaliyadda, H.M.V.R. Herath, and M.P.B. Ekanayake are with the Department of Electrical and Electronic Engineering, University of Peradeniya, KY 20400, Sri Lanka (email: kavingaweerasooriya@eng.pdn.ac.lk, e14273@eng.pdn.ac.lk, sanjaya.h@eng.pdn.ac.lk, roshangodd@ee.pdn.ac.lk, vijitha@eng.pdn.ac.lk; mpb.ekanayake@ee.pdn.ac.lk).}%
\thanks{B. Rathnayake is with the Department of Electrical, Computer, and Systems Engineering, Rensselaer Polytechnic Institute, Troy, NY 12180, USA (email: rathnb@rpi.edu).}%
}
\maketitle
\begin{abstract}
Hyperspectral unmixing (HU) has become an important technique in exploiting hyperspectral data since it decomposes a mixed pixel into a collection of endmembers weighted by fractional abundances. The endmembers of a hyperspectral image (HSI) are more likely to be generated by independent sources and be mixed in a macroscopic degree before arriving at the sensor element of the imaging spectrometer as mixed spectra. Over the past few decades, many attempts have focused on imposing auxiliary regularizes on the conventional nonnegative matrix factorization (NMF) framework in order to effectively unmix these mixed spectra. As a promising step toward finding an optimum regularizer to extract endmembers, this paper presents a novel blind HU algorithm, referred to as Kurtosis-based Smooth Nonnegative Matrix Factorization (KbSNMF) which incorporates a novel regularizer based on the statistical independence of the probability density functions of endmember spectra. Imposing this regularizer on the conventional NMF framework promotes the extraction of independent endmembers while further enhancing the parts-based representation of data. Experiments conducted on diverse synthetic HSI datasets (with numerous numbers of endmembers, spectral bands, pixels, and noise levels) and three standard real HSI datasets demonstrate the validity of the proposed KbSNMF algorithm compared to several state-of-the-art NMF-based HU baselines. The proposed algorithm exhibits superior performance especially in terms of extracting endmember spectra from hyperspectral data; therefore, it could
uplift the performance of recent deep learning HU methods which
utilize the endmember spectra as supervisory input data for
abundance extraction.
\end{abstract}


\newpage
\section{Introduction} 
\label{section:Introduction}
\IEEEPARstart{H}{yperspectral} image (HSI) technology has become a leading imaging technology in many fields including medical imaging, food quality assessment, forensic sciences, surveillance, and remote sensing \cite{8314827}. However, due to the insufficient spatial resolution of spectrometers and homogeneous mixture of distinct macroscopic materials in imaging scenes, the observed reflectance spectrum at each pixel of an HSI could easily be a mixture of spectra belonging to a set of constituent members (also called endmembers). This mixing phenomenon constitutes a major concern with regard to many applications. As a remedy to this complication, various methods of hyperspectral unmixing (HU) have been implemented to extract endmember spectra along with their fractional composition (also called abundances). HU is a study of three subproblems, \textit{i.e.} determining the no. of endmembers, extracting the endmember spectra, and realizing their abundances \cite{keshava2002spectral}.

In the past, many algorithms have been introduced in order to solve the HU problem \cite{qin2020blind, 4694061, 7173001, 6923488, lu2013manifold, 5871318, halimi2011nonlinear,7194802, 9186270, yokoya2012coupled, iordache2011sparse, 7892882, 8616834} and these algorithms can be categorized under three main schemes according to the basic computational approaches \cite{6200362}: 1- statistical algorithms, 2- geometric algorithms and 3- sparse regression based unmxing algorithms. Statistical algorithms interpret a mixed pixel by utilizing statistical representations. These representations are commonly analytical expressions based on the probability density functions of the underlying mixed pixel spectra. Bayesian self organizing maps (BSOM) \cite{4423225}, independent component analysis (ICA) \cite{10.5555/983149, 1677768}, independent factor analysis (IFA) \cite{1381633}, dependent component analysis (DECA) \cite{4423734}, automated morphological endmember extraction (AMEE) \cite{1046852}, and spatial-spectral endmember extraction algorithm (SSEE) \cite{ROGGE2007287} are some of the popular statistical algorithms utilized for HU. Geometric algorithms exploit the geometric orientation of HSI data in an $n$-dimensional space, where $n$ is the no. of spectral bands captured by the imaging spectrometer. Vertex component analysis (VCA) \cite{1411995}, minimum volume transform (MVT) \cite{297973}, simplex identification via split augmented Lagrangian (SISAL) \cite{5289072}, optical real-time adaptive spectral identification system (ORASIS) \cite{10.1117/12.221352}, iterative error analysis (IEA) \cite{canada}, and nonnegative matrix factorization (NMF) \cite{2ef7006f34ff4cd7afa86c9bc8932c80} are some of the geometric algorithms frequently utilized for HU. Sparse regression based approaches utilize known libraries.  The unmixing problem is formulated as a sparse linear regression problem which is based on the assumption that every feature can be linearly created by few elements extracted from known libraries \cite{5692827, 6471206, 6568877, 6675070}.  

In the recent past, several approaches have been introduced where deep learning (DL) is utilized for HU. More often, DL-based methods for HU do not perform blind unmixing, \textit{i.e.} extract both endmember spectra and abundances. In \cite{7473851}, a Hopfield neural network (HNN) machine learning approach is utilized to solve the seminonnegative matrix factorization problem, which has illustrated promising performance with regard to abundance extraction when given reliable endmember spectra as supervisory input data. In \cite{7120463}, an artificial neural network (ANN) is utilized to inverse the pixel spectral mixture in Landsat imagery. Here, to train the network, a spectral library had been created, consisting of endmember spectra collected from the image and simulated mixed spectra. In \cite{5967899}, a two-staged ANN architecture has been introduced in which the first stage reduces the dimension of the input vector utilizing endmember spectra as input data. As can be seen, most of the current DL-based methods for HU utilize endmember spectra as supervisory input data in order to extract the abundances.

Originally introduced by Lee and Seung \cite{2ef7006f34ff4cd7afa86c9bc8932c80}, NMF is a mathematical tool which is utilized to decompose a nonnegative data matrix into the product of two other nonnegative matrices of lower rank based on the optimization of a particular objective function. Since the nonnegativity criterion does not accommodate any negative elements in resultant matrices \cite{Lee1999}, which also coincides with the objective of HU. Driven by this parts-based representation of the NMF framework, NMF-based algorithms are often utilized to solve the HU problem. However, NMF is an ill-posed geometric algorithm; therefore, it does not possess a unique solution \cite{5871318}. The non-convex objective functions utilized for NMF compel its solution space to be wide. Thus, many researchers have introduced novel NMF algorithms by adding different auxiliary regularizes to the conventional NMF framework in order to improve the uniqueness of its solution with respect to the HU setting. $l_{1/2}$-sparsity constrained NMF ($l_{1/2}$-NMF) \cite{5871318}, spatial group sparsity regularized NMF (SGSNMF) \cite{7995123}, minimum volume rank deficient NMF (Min-vol NMF) \cite{8682280}, manifold regularized sparse NMF \cite{lu2013manifold}, Double Constrained NMF \cite{lu2014double}, total variation regularized reweighted sparse NMF (TV-RSNMF) \cite{he2017total}, subspace clustering constrained sparse NMF (SC-NMF) \cite{ lu2020subspace}, nonsmooth NMF (\textit{ns}NMF) \cite{1580485}, robust collaborative NMF (R-CoNMF) \cite{7505959}, Subspace  Structure  Regularized NMF (SSRNMF) \cite{9146211}, graph regularized NMF (GNMF) \cite{5674058} and Projection-Based NMF (PNMF) \cite{7120079} are some customary NMF-based baselines utilized for HU. Furthermore, A new architecture has recently emerged for blind unmixing under the premise Nonnegative Tensor factorization (NTF). In the paper, Matrix-Vector NTF for Blind Unmixing of Hyperspectral Imagery (MVNTF) \cite{7784711}, the authors have formalized a novel way of unmixing while preserving the spatial information by factorizing hyper spectral 3D cubes instead of unwrapped spectral datasets.   

In HU, the endmembers are typically macroscopic objects in the HSI scene, such as soil, water, vegetation, etc \cite{keshava2002spectral}. In a broader sense, HU attempts to find these macroscopic objects by utilizing the observations of signals that have already interacted (or mixed) with other objects in the scene before arriving at the sensing element of the imaging spectrometer. It is pragmatic to assume that the endmembers are consequences of different physical processes; hence, statistically independent\footnote{Throughout the rest of this paper, we refer to the ``statistical independence'' of endmembers as the ``independence'' of endmembers.} \cite{10.5555/983149}. If a particular methodology promotes maximizing the independence of endmembers, each of the endmember spectra extracted utilizing that particular method will be more independent than the mixed pixel spectra. Therefore, such a method would be a progression toward the extraction of more realistic endmember spectra belonging to independent macroscopic objects. Even though the frequently-associated abundance sum-to-one (ASC) constraint \cite{911111} in HU does not accommodate the concept of independent endmembers, algorithms such as ICA \cite{10.5555/983149, 1677768}, IFA \cite{1381633}, and independent innovation analysis (IIA) \cite{4667135} are popular algorithms utilized in HU which consider this concept. Also, several attempts have been taken previously in order to incorporate the independence of endmembers onto the conventional NMF framework. The authors of \cite{kitamura2016determined} have proposed a novel initialization method based on statistical independence between NMF components. In \cite{6648948}, an attempt has been made to initialize NMF with a modified ICA method (modifICA-NMF). In \cite{7486081}, a novel effective method has been introduced unifying independent vector analysis (IVA) and NMF. Our previous work \cite{8899045} discusses the suitability of utilizing the fundamental notions of kurtosis-based ICA to enhance the conventional NMF algorithm.

Inspired by the interpretable parts-based representations and simplicity of imposing auxiliary regularizes of the conventional NMF framework, and motivated by our previous work \cite{8899045,9033972,8747025,jsi.2019.a11,jnsf, 9063280, 9063263}, this study proposes a novel regularizer to the conventional NMF framework named Average Kurtosis regularizer. Incorporating this regularizer along with an abundance smoothing mechanism, we present a novel blind HU algorithm named Kurtosis-based Smooth Nonnegative Matrix Factorization (KbSNMF) along with its two variants KbSNMF-fnorm and KbSNMF-div. The motivation of the proposed work is to promote the independence of endmembers while extracting them in accordance with the parts-based representations of the conventional NMF framework, thereby attempting to extract the most realistic endmember spectra from a given HSI. The contributions of this paper are summarized as follows:
\begin{enumerate}
\item Introduction of a novel regularizer for HU, based on kurtosis, which promotes the independence of endmembers of an HSI.
\item Computation of the gradient of the aforesaid regularizer w.r.t. the factors of the conventional NMF framework, and the establishment of a blind HU algorithm named KbSNMF, which effectively promotes the independence of endmembers while maintaining the smoothness of abundance maps. 

\end{enumerate}

We also implement and evaluate the performance of the proposed algorithm in comparison with several selected state-of-the-art NMF-based HU baselines. Experiments are conducted on diverse synthetic HSI datasets (with numerous numbers of endmembers, spectral bands, pixels, and noise levels) as well as on three standard real HSI datasets. These experiments substantiate that the proposed algorithm outperforms other state-of-the-art NMF-based blind HU algorithms in many instances, especially in extracting endmember spectra. This observation is understandable since the proposed algorithm tries to improve upon the pragmatic characteristics of the endmember spectra, rather than trying to improve upon the pragmatic characteristics of the abundance maps. Thus, in an unsupervised setting where there is the luxury of utilizing a DL-based method for abundance extraction, the proposed algorithm would provide a viable counterpart to generate endmember spectra as supervisory input data to the DL-based method.	

The rest of the paper is arranged as follows. Section \ref{section:Background} provides the background related to the proposed algorithm. In Section \ref{section:Average Kurtosis Regularizer}, the novel kurtosis-based regularizer is developed along with its derivatives. In Section \ref{section:Kurtosis-based Smooth Nonnegative Matrix Factorization (KbSNMF)}, the novel KbSNMF algorithm is introduced. Section \ref{section:Algorithm Implementation} discusses some key issues related to the implementation of the proposed algorithm. Section \ref{section:Experiments and Discussions} is devoted for experimental results and the paper is concluded in Section \ref{section:Conclusion}.

\section{Background}
\label{section:Background}

\subsection{Linear Mixture Model (LMM)}
\label{section:Linear Mixture Model (LMM)}

LMM is the most frequently utilized model for HU and its implications had been widely discussed in many previous works \cite{iordache2012total, 6923488, 5871318, 1411995, 4694061}. This model highly depends on the assumption that the incident light waves reflect only once from the underlying macroscopic objects and are captured by the sensing element of the imaging spectrometer without being subjected to scattering. In the LMM, the spectrum at each pixel is represented as a linear combination of the endmember spectra as below,
\begin{equation} 
\label{eqLMM1}
\mathbf{x}_{j} = \sum_{i=1}^{r} S_{ij}\mathbf{a}_{i} + \mathbf{e}_{j}
\end{equation}
where $\mathbf{x}_{j}\in\mathbb{R}_{+}^{n\times 1}$ is the $j^{\text{th}}$ pixel spectrum, $S_{ij}$ is the fractional composition occupied by the $i^{\text{th}}$ endmember in the $j^{\text{th}}$ pixel, $\mathbf{a}_{i}\in\mathbb{R}_{+}^{n\times 1}$ is the spectrum of the $i^{\text{th}}$ endmember of the HSI, $\mathbf{e}_{j}\in\mathbb{R}^{n\times 1}$ is an additive Gaussian noise associated with modeling errors, and $r$ is the no. of endmembers in the HSI. All spectra are measured in reflectance values; hence, the nonnegativity in $\mathbf{x}_{j}$'s and $\mathbf{a}_{i}$'s. The nonnegativity constraint $S_{ij}\geq0$ and the sum-to-one constraint $\sum_{i=1}^{r} S_{ij}=1$ are implied in order to guarantee that the fractional compositions representing the endmembers are nonnegative and the abundance summation equals 1 at each pixel. The LMM can be reformulated in matrix notations as below,
\begin{equation} 
\label{eqLMM2}
\mathbf{X} = \mathbf{A}\times \mathbf{S} + \mathbf{E}
\end{equation}
where $\mathbf{X}\in\mathbb{R}_{+}^{n\times m}$ is the HSI data matrix, $n$ being the no. of spectral bands and $m$ being the no. of pixels of the HSI, $\mathbf{A}\in\mathbb{R}_{+}^{n\times r}$ is the endmember matrix whose columns represent the spectra of each of the $r$ endmembers, $\mathbf{S}\in\mathbb{R}_{+}^{r\times m}$ is the abundance matrix whose columns represent the fractional compositions at each of the $m$ pixels, and $\mathbf{E}\in\mathbb{R}^{n\times m}$ is the noise matrix. This formulation casts the HU problem as a BSS problem, \textit{i.e.} simultaneous extraction of the endmember spectra and their abundances at each pixel while utilizing the HSI as the input.

\begin{figure*}
\centering
\includegraphics[width=\columnwidth*2]{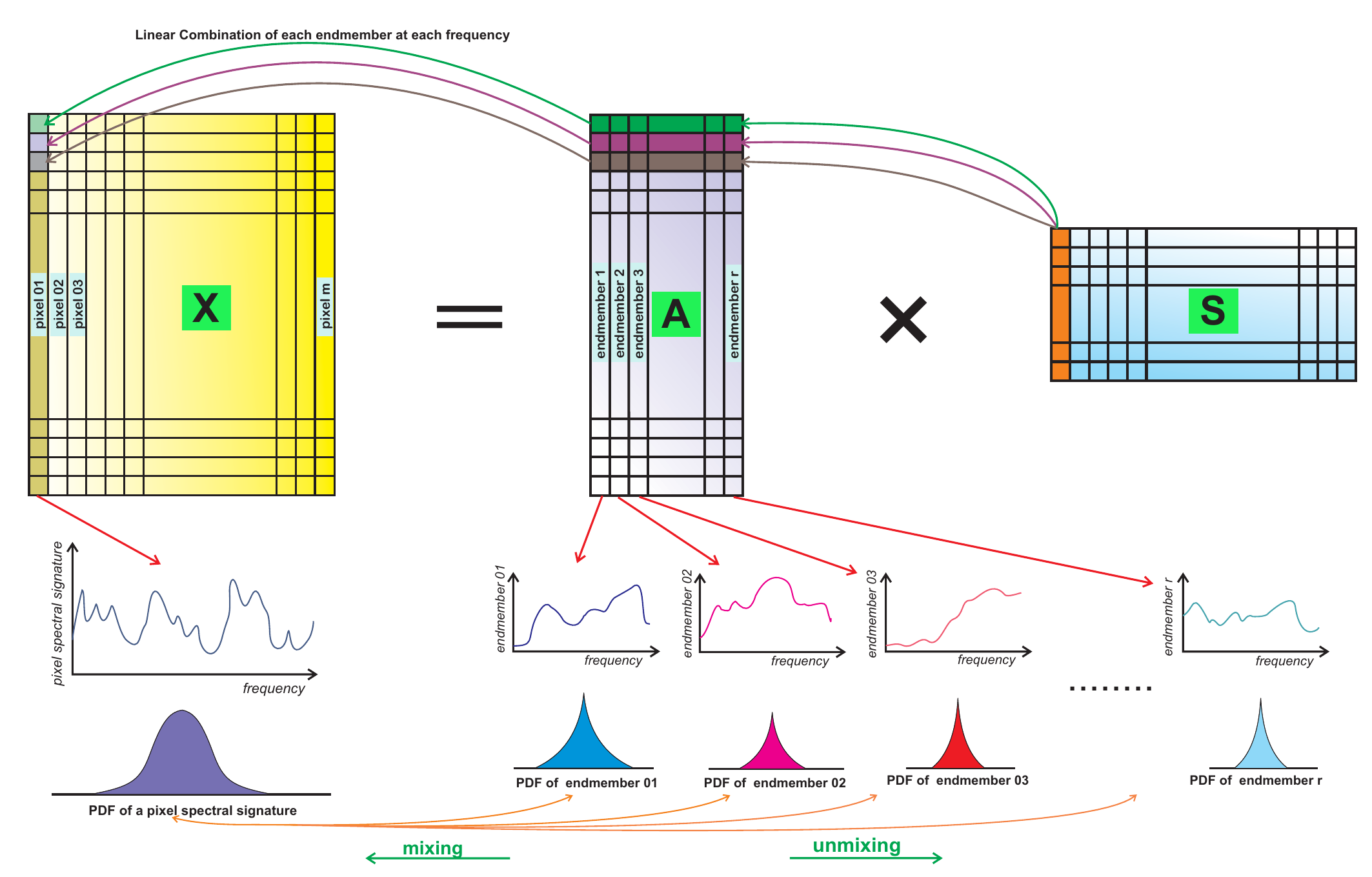}
\caption{Underlying mechanism of the proposed algorithm. According to the unmixing strategy, it is discernible that every pixel is  a linear combination of several independent endmembers. The proposed method promotes independence of endmembers by increasing the super-gaussianity. The algorithm concept is a derivative of the central limit theorem.}
\label{gra_abstract}
\end{figure*}

\subsection{Nonnegative Matrix Factorization (NMF)}
\label{section:Nonnegative Matrix Factorization (NMF)}

NMF is a low-rank approximation of nonnegative matrices widely utilized in the fields of computer vision, clustering, data compression, etc \cite{kitamura2016determined,weixiang2006nonnegative,textclassi,berry2007algorithms, 5593218}. NMF was first introduced by Lee and Seung \cite{2ef7006f34ff4cd7afa86c9bc8932c80} as a parts-based representation technique which permits the data in a nonnegative matrix to be decomposed into two other nonnegative matrices. Given a matrix $\mathbf{V}\in\mathbb{R}_{+}^{n\times m}$, NMF tries to find nonnegative matrices $\mathbf{W}\in\mathbb{R}_{+}^{n\times r}$ (known as the source matrix) and $\mathbf{H}\in\mathbb{R}_{+}^{r\times m}$ (known as the mixing matrix) which satisfy the approximation below.
\begin{equation} 
\label{eqNMF1}
\mathbf{V}\approx \mathbf{WH} 	
\end{equation}
However, there are infinite no. of $\mathbf{W}$, $\mathbf{H}$ solution pairs which satisfy the above approximation. For instance, it is possible to write $\mathbf{WH} = (\mathbf{W\Gamma}^{-1})(\mathbf{\Gamma H})$ for any invertible $\mathbf{\Gamma}\in\mathbb{R}_{+}^{r\times r}$. The conventional procedure to achieve (\ref{eqNMF1}) is by defining an objective function which quantifies the quality of the approximation between $\mathbf{V}$ and $\mathbf{WH}$ and implementing an optimization algorithm to minimize the defined objective function w.r.t. $\mathbf{W}$ and $\mathbf{H}$ . One of the most commonly utilized objective function is the square of the Frobenius norm between $\mathbf{V}$ and $\mathbf{WH}$ as in (\ref{dist1}).
\begin{equation} 
\label{dist1}
\Vert\mathbf{V}-\mathbf{WH}\Vert^{2}_{F} = \sum_{ij}\left(V_{ij}-(WH)_{ij}\right)^{2}
\end{equation}
The above expression is lower bounded by zero and distinctly vanishes if and only if $\mathbf{V} = \mathbf{WH}$. Another popular objective function is the divergence\footnote{Unlike the Frobenius norm, the divergence cannot be designated as a ``distance'' since it is not symmetric in $\mathbf{V}$ and $\mathbf{WH}$. Thus, it is the common practice to refer to it as the ``divergence of $\mathbf{V}$ from $\mathbf{WH}$''} of $\mathbf{V}$ from $\mathbf{WH}$ as in (\ref{dist2}).   
\begin{equation} 
\label{dist2}
D(\mathbf{V}\Vert\mathbf{WH}) = \sum_{ij}\left(V_{ij}\log{\frac{V_{ij}}{(WH)_{ij}}}-V_{ij}+(WH)_{ij}\right)
\end{equation}
Similar to the Frobenius norm, the divergence is also lower bounded by zero and vanishes if and only if $\mathbf{V} = \mathbf{WH}$. Even though (\ref{dist1}) and (\ref{dist2}) functions are convex in $\mathbf{W}$ and $\mathbf{H}$ alone, they are not convex in $\mathbf{W}$ and $\mathbf{H}$ together \cite{2ef7006f34ff4cd7afa86c9bc8932c80}. Hence, it is not possible to analytically find global minima of these functions w.r.t. $\mathbf{W}$ and $\mathbf{H}$. However, it is possible to find local minima utilizing numerical optimization methods. Lee and Seung \cite{2ef7006f34ff4cd7afa86c9bc8932c80} have proposed the below (\ref{update1_W_H}) and (\ref{update2_W_H}) multiplicative update rules to find local minima of the above (\ref{dist1}) and (\ref{dist2}) functions respectively.
\begin{equation} 
\label{update1_W_H}
\mathbf{W}\leftarrow\mathbf{W}\circ\frac{\mathbf{VH}^{T}}{\mathbf{WHH}^{T}},\qquad
\mathbf{H}\leftarrow \mathbf{H}\circ\frac{\mathbf{W}^{T}\mathbf{V}}{\mathbf{W}^{T}\mathbf{WH}}
\end{equation} 
\begin{equation} 
\label{update2_W_H}
\mathbf{W}\leftarrow\mathbf{W}\circ\frac{\frac{\mathbf{V}}{\mathbf{WH}}\mathbf{H}^{T}}{\mathbf{1}_{n\times m}\mathbf{H}^{T}},\qquad
\mathbf{H}\leftarrow \mathbf{H}\circ\frac{\mathbf{W}^{T}{\frac{\mathbf{V}}{\mathbf{WH}}}}{\mathbf{W}^{T}\mathbf{1}_{n\times m}}
\end{equation} 
Lee and Seung have further proven the convergence of both the above update rules utilizing an auxiliary function analogous to the proof of convergence of the Expectation Maximization algorithm \cite{2ef7006f34ff4cd7afa86c9bc8932c80}.

The LMM model transforms the HU problem into the form of a conventional NMF problem. If $\mathbf{V}$ is the HSI data matrix $\mathbf{X}$, then source matrix $\mathbf{W}$ is the endmember matrix $\mathbf{A}$ and mixing matrix $\mathbf{H}$ is the abundance matrix $\mathbf{S}$. Thus, given $\mathbf{X}$, solving the blind HU problem for $\mathbf{A}$ and $\mathbf{S}$ utilizing the conventional NMF problem can be formulated as in (\ref{mindist1}) and (\ref{mindist2}) for Frobenius norm and divergence-based objective functions respectively.
\begin{equation} 
\label{mindist1}
\argmin_{\mathbf{A},\mathbf{S}}\Vert\mathbf{X}-\mathbf{AS}\Vert^{2}_{F}, \hspace{10pt} \text{s.t. } \mathbf{A},\mathbf{S}\succeq 0
\end{equation}
\begin{equation} 
\label{mindist2}
\argmin_{\mathbf{A},\mathbf{S}}D(\mathbf{X}\Vert\mathbf{AS}), \hspace{10pt} \text{s.t. }\mathbf{A},\mathbf{S}\succeq 0 
\end{equation}
In order to solve above problems while improving the uniqueness, many previous works have incorporated additional auxiliary regularizes on $\mathbf{A}$ and $\mathbf{S}$ \cite{8682280,7505959,5871318,7995123,5674058, 7194802}.   

\section{Average Kurtosis Regularizer} 
\label{section:Average Kurtosis Regularizer}

\subsection{Kurtosis of a Signal}
\label{section:Kurtosis of a Signal}

Central moments are often utilized in signal processing in order to characterize the spread of the probability density function (pdf) of a signal \cite{10.5555/983149}. A normalized version of the fourth central moment, given by (\ref{kurtosis}), is called the kurtosis of a signal. Here $y$ denotes the signal, $\overline{y}$ denotes the mean of the signal, and $\mathbb{E}$ is the expectation operator. Intuitively, kurtosis provides a measure of the ``peaky''ness of the shape of the pdf of a signal. Excess kurtosis is a measure which compares the kurtosis of a given pdf with the kurtosis of a Gaussian distribution. Since the kurtosis of a Gaussian distribution equals 3, the excess kurtosis can be defined as in (\ref{excesskurt}).
\begin{equation} 
\label{kurtosis}
\text{kurtosis} = \frac{\mathbb{E}[(y-\overline{y})^{4}]}{(\mathbb{E}[(y-\overline{y})^{2}])^2}
\end{equation}
\begin{equation} 
\label{excesskurt}
\text{excess kurtosis} = \text{kurtosis} - 3
\end{equation}
Based on the value of excess kurtosis, distributions are categorized under three main types. \textit{Mesokurtic} distribution is close to a Gaussian distribution; has an excess kurtosis closer to zero. \textit{Leptokurtic} (also known as super-Gaussian) distribution has a higher and sharper central peak; tails are longer and flatter; has positive excess kurtosis. \textit{Platykurtic} (also known as sub-Gaussian) distribution has a lower and broader central peak; tails are shorter and thinner; has negative excess kurtosis.

Central Limit Theorem (CLT) ensures that a mixture of signals is approximately Gaussian irrespective of the distributions of the underlying source signals. Even though the converse of CLT is not assured, \textit{i.e.} it is not certain that any Gaussian signal is a mixture of non-Gaussian signals, in practical scenarios, Gaussian signals do consist of a mixture of non-Gaussian signals \cite{10.5555/983149}. Thus, to extract the underlying source signals from a signal mixture, it is common practice in BSS to define a measure of non-Gaussianity and implement an algorithm which maximizes the defined measure as Fig. \ref{gra_abstract} illustrates. Subsequently, excess kurtosis seems to be a suitable candidate for this purpose as it is a measure of non-Gaussianity. If the excess kurtosis value of a signal is close to zero, it tempts to be Gaussian and if the excess kurtosis value of a signal is away from zero, it tempts to be non-Gaussian (super- or sub-Gaussian). Since there are two types of non-Gaussian distributions,  it is common practice in most BSS methods to assume that source signals are of only one type. In this work, we assume the constituent spectra of an HSI to have super-Gaussian distributions. Hence, from a given HSI data matrix $\mathbf{X}$, we aim to extract an endmember matrix $\mathbf{A}$, whose column-wise average kurtosis is maximized, utilizing an NMF framework. Thus, we introduce a novel constrained NMF algorithm which incorporates the maximization of the average kurtosis of endmembers.

\begin{figure*}[!t]
\centering
\includegraphics[width=\columnwidth*2]{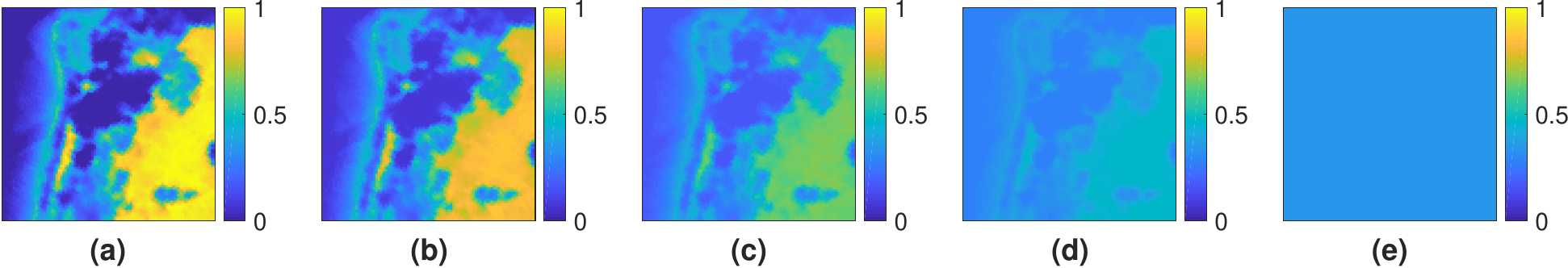}
\caption{Effects of the smoothing parameter demonstrated on a ground truth abundance map (“Soil”) of a real HSI dataset (“Samson”): (a) $\theta $= 0  (no smoothing), (b) $\theta $ = 0.2, (c) $\theta $ = 0.5, (d) $\theta $ = 0.8, (e) $\theta $ = 1 (maximum smoothing).}
\label{smoothing}
\end{figure*}

\subsection{Average Kurtosis} 
\label{section:Average Kurtosis}

Obeying the notations introduced in Section \ref{section:Linear Mixture Model (LMM)}, $\mathbf{A}\in\mathbb{R}_{+}^{n\times r}$ is the endmember matrix whose columns represent the spectra of each of the $r$ endmembers of the HSI. Thus, it is possible to extract the $i^{\text{th}}$ endmember utilizing a simple matrix manipulation as below,
\begin{equation} 
\label{endmember}
\mathbf{a}_{i} = \mathbf{A}\mathbf{\Phi}_{i}
\end{equation}
where $\mathbf{a}_{i}$ is spectrum of the $i^{\text{th}}$ endmember from matrix $\mathbf{A}$ (or the  $i^{\text{th}}$ column of matrix $\mathbf{A}$) and $\mathbf{\Phi}_{i}\in\mathbb{R}^{r\times 1}$ is a column vector whose all elements are zeros except for the $i^{\text{th}}$ element which equals 1. If the kurtosis of the $i^{\text{th}}$ endmember is ${K}_{i}$, it can expressed as below according to (\ref{kurtosis}).
\begin{equation} 
\label{Kurt_end}
{K}_{i} = \frac{\mathbb{E}[(\mathbf{a}_{i}-\overline{\mathbf{a}_{i}})^{4}]}{(\mathbb{E}[(\mathbf{a}_{i}-\overline{\mathbf{a}_{i}})^{2}])^2}
\end{equation}
Thus, the average kurtosis through all $r$ endmembers, $\overline{K}$ can be expressed as below utilizing (\ref{endmember}) and (\ref{Kurt_end}).
\begin{equation} 
\label{avg_kurt}
\begin{split}
\overline{K} & = \frac{1}{r}\sum_{q=1}^{r}{K}_{q} \\
& =
\frac{1}{r}\sum_{q=1}^{r}\frac{\mathbb{E}[(\mathbf{A}\mathbf{\Phi}_{q}-\overline{\mathbf{A}\mathbf{\Phi}_{q}})^{4}]}{(\mathbb{E}[(\mathbf{A}\mathbf{\Phi}_{q}-\overline{\mathbf{A}\mathbf{\Phi}_{q}})^{2}])^2}
\end{split}
\end{equation}
Thus, it is seen that $\overline{K}$ is a function of $\mathbf{A}$; therefore, it can be written as $\overline{K}(\mathbf{A})$. We try to maximize $\overline{K}$ so that the extracted endmembers will have a higher average kurtosis, \textit{i.e.} they will be closer to super-Gaussian signals. Hence, the proposed framework would favorably influence the extraction of more realistic endmember spectra from the underlying HSI.

\subsection{Derivative of Average Kurtosis} 
\label{section:Derivative of Average Kurtosis}

In order to incorporate the average kurtosis regularizer onto the conventional NMF framework, it is essential to find the gradient (or the partial derivative) of $\overline{K}$ w.r.t $\mathbf{A}$ and $\mathbf{S}$, \textit{i.e.} $\nabla_{\mathbf{A}}\overline{K}\in\mathbb{R}^{n\times r}$ and $\nabla_{\mathbf{S}}\overline{K}\in\mathbb{R}^{r\times m}$. Since $\overline{K}$ is not a function of $\mathbf{S}$, $\nabla_{\mathbf{S}}\overline{K} = \mathbf{0}\in\mathbb{R}^{n\times r}$.  In this section, we provide a detailed explanation on finding $\nabla_{\mathbf{A}}\overline{K}$. Since $\mathbf{A}$ is the endmember matrix, we denote each of its elements by the notation $A_{ki}$, with the meaning of the reflectance value belonging to the $k^{\text{th}}$ spectral band of the $i^{\text{th}}$ endmember. Thus, the $(k,i)^{\text{th}}$ element of $\nabla_{\mathbf{A}}\overline{K}$ can be written as below implementing an element-wise derivative.
\begin{equation}
\label{derivation1}
\begin{split}
\nabla_{\mathbf{A}}\overline{K}_{ki} & = \frac{\partial \overline{K}}{\partial A_{ki}}\\
& = \frac{1}{r}\sum_{q=1}^{r}\frac{\partial K_{q}}{\partial A_{ki}}
\end{split}
\end{equation}
where
\begin{equation}\label{derivation2}
\frac{\partial K_{q}}{\partial A_{ki}} = \begin{cases}
\frac{\partial K_{i}}{\partial A_{ki}},& \text{if } q = i\\
0. & \text{otherwise}
\end{cases}\
\end{equation}
For the convenience of simplifying, we assume that each of the endmember spectra vectors have unit variance, \textit{i.e.} $(\mathbb{E}[(\mathbf{a}_{i}-\overline{\mathbf{a}_{i}})^{2}])^2 = 1, \enskip \forall i$. In order to rectify the effects of this assumption, a normalization step is carried out as discussed in Section \ref{section:Normalization}. As a result, we obtain a simplified version of $\nabla_{\mathbf{A}}\overline{K}_{ki}$ as below,
\begin{equation} 
\label{derivation3}
\begin{split}
\nabla_{\mathbf{A}}\overline{K}_{ki} & = \frac{1}{r}\frac{\partial \left[\mathbb{E}[(\mathbf{a}_{i}-\overline{\mathbf{a}_{i}})^{4}]\right]}{\partial A_{ki}}\\
& = \frac{1}{nr}{\sum_{p=1}^{n}\frac{\partial(A_{pi}-\mu_{i})^{4}}{\partial A_{ki}}}\\
\end{split}
\end{equation}
where $A_{pi}$ is the reflectance value belonging to the $p^{\text{th}}$ spectral band of the $i^{\text{th}}$ endmember, and $\mu_{i}$ is the mean reflectance of the $i^{\text{th}}$ endmember. As can be seen, $\nabla_{\mathbf{A}}\overline{K}_{ki}$ is a summation of $n$ more partial derivative terms for which the solutions can be obtained by utilizing the chain rule in calculus. 
\begin{equation}\label{derivation4}
\frac{\partial}{\partial A_{ki}}(A_{pi}-\mu_{i})^{4} = \begin{cases}
-4(A_{ki}-\mu_{i})^{3} \left(\frac{1}{n}-1\right),& \text{if } p = k\\
-4 (A_{pi}-\mu_{i})^{3} \left(\frac{1}{n}\right). & \text{otherwise}
\end{cases}\
\end{equation}
Thus, the partial derivative term  $\nabla_{\mathbf{A}}\overline{K}_{ki}$ in (\ref{derivation3}) can be written as follows.
\begin{equation} 
\label{derivation5}
\nabla_{\mathbf{A}}\overline{K}_{ki} =\frac{-4}{nr}\left[S_{i}-(A_{ki}-\mu_{i})^{3}\right]
\end{equation}
where $S_{i}=\frac{1}{n}\sum_{p=1}^{n}(A_{pi}-\mu_{i})^{3}$ represents a normalized version of the third central moment (skewness) of the $i^{\text{th}}$ endmember. However, in this work we do not explore the implication of skewness within derivative of the average kurtosis. Concatenating the element-wise derivatives, we then express $\nabla_{\mathbf{A}}\overline{K}$ as the difference between two matrices as below,
\begin{equation}
\label{derivation6}
\nabla_{\mathbf{A}}\overline{K} = \frac{-4}{nr}\left(\mathbf{P}-\mathbf{Q}\right)
\end{equation}
where $P_{ki} = S_{i}$ and $Q_{ki} = (A_{ki}-\mu_{i})^{3}$. Then, $\mathbf{Q}$ and $\mathbf{P}$ can be written as in (\ref{derivation7}) and (\ref{derivation8}) respectively for the convenience of incorporating $\nabla_{\mathbf{A}}\overline{K}$ in the NMF framework. 
\begin{equation}
\label{derivation7}
\mathbf{Q} = \left[\mathbf{A}-\frac{1}{n}\mathbf{1}_{n\times n}\mathbf{A}\right]^{\circ 3}
= \left[\mathbf{N}\mathbf{A}\right]^{\circ 3}
\end{equation}
\begin{equation}
\label{derivation8}
\mathbf{P} =\frac{1}{n}\mathbf{1}_{n\times n}\mathbf{Q}\\
= \frac{1}{n}\mathbf{1}_{n\times n}\left[\mathbf{N}\mathbf{A}\right]^{\circ 3}
\end{equation}
where $\mathbf{N}=(\mathbf{I}-\frac{1}{n}\mathbf{1}_{n\times n})$ and $\mathbf{1} \in\mathbb{R}^{n\times n}$ denotes a matrix whose all elements are ones. $[.]^{\circ 3}$ denotes the Hadamard (element-by-element) power by 3. Finally, from (\ref{derivation6}), $\nabla_{\mathbf{A}}\overline{K}$ can be written as follows.  
\begin{equation}
\label{derivation9}
\begin{split}
\nabla_{\mathbf{A}}\overline{K}&=\frac{-4}{nr}\left[\frac{1}{n}\mathbf{1}_{n\times n}\left[\mathbf{N}\mathbf{A}\right]^{\circ 3}-\left[\mathbf{N}\mathbf{A}\right]^{\circ 3}\right]\\
& = \frac{4}{nr}\left[\mathbf{N}\left[\mathbf{N}\mathbf{A}\right]^{\circ 3}\right]\\
\end{split}
\end{equation}

\section{Kurtosis-based Smooth Nonnegative Matrix Factorization (KbSNMF)} 
\label{section:Kurtosis-based Smooth Nonnegative Matrix Factorization (KbSNMF)}

In this section, we propose a novel blind HU algorithm which not only promotes the independence of endmembers via the kurtosis regularizer but also promotes the smoothness of the abundance maps by integrating a smoothing matrix to the conventional NMF framework. Hence, we denominate the proposed algorithm as Kurtosis-based Smooth NMF (KbSNMF). In the proceeding sections, we discuss two variants of KbSNMF depending on the objective function utilized for approximation.

\subsection{KbSNMF-fnorm} 
\label{section:KbSNMF-fnorm}

Here we present KbSNMF based on Frobenius norm (KbSNMF-fnorm). The general optimization problem for KbSNMF-fnorm is as below.
\begin{equation} 
\label{objectnew1}
\argmin_{\mathbf{A},\mathbf{S}}\left\{\Vert\mathbf{X}-\mathbf{AMS}\Vert^{2}_{F} - \gamma \overline{K}(\mathbf{A})\right\}, \hspace{10pt} s.t.\mathbf{A},\mathbf{S}\succeq 0
\end{equation}
Here,  $\gamma\in\mathbb{R}_{+}$ is a parameter which establishes the trade-off between approximation error and non-Gaussianity of the endmembers rendered by $\overline{K}$, and $\mathbf{M}\in\mathbb{R}_{+}^{r\times r}$ is a symmetric matrix called the smoothing matrix which is defined as below,
\begin{equation} 
\label{smoothing_matrix}
\mathbf{M} = (1-\theta)\mathbf{I}+\frac{\theta}{r}\mathbf{1}_{r\times 1}\mathbf{1}_{r\times 1}^{T}
\end{equation}
where $\mathbf{I}$ is the identity matrix, $\mathbf{1}\in\mathbb{R}^{r\times 1}$ is a vector whose all elements are ones, and $\theta$ is a parameter which satisfies $0 \leq \theta \leq 1$ and controls the extent of smoothness. Enforcing smoothness onto the abundance matrix can be interpreted as $\mathbf{Y = MS}$, where $\mathbf{Y}$ is the smoothness-enforced abundance matrix. When $\theta = 0$, $\mathbf{M} = \mathbf{I}$, hence, $\mathbf{Y = S}$ and no smoothing has occurred in $\mathbf{S}$. As $\theta\rightarrow1$, $\mathbf{Y}$ tends to become smoother and reaches the smoothest possible at $\theta = 1$. Fig. \ref{smoothing} demonstrates the effects of smoothing parameter $\theta$ on the abundance maps.
\begin{figure*}[!b]
\centering
\includegraphics[width=\columnwidth*2,height=8cm]{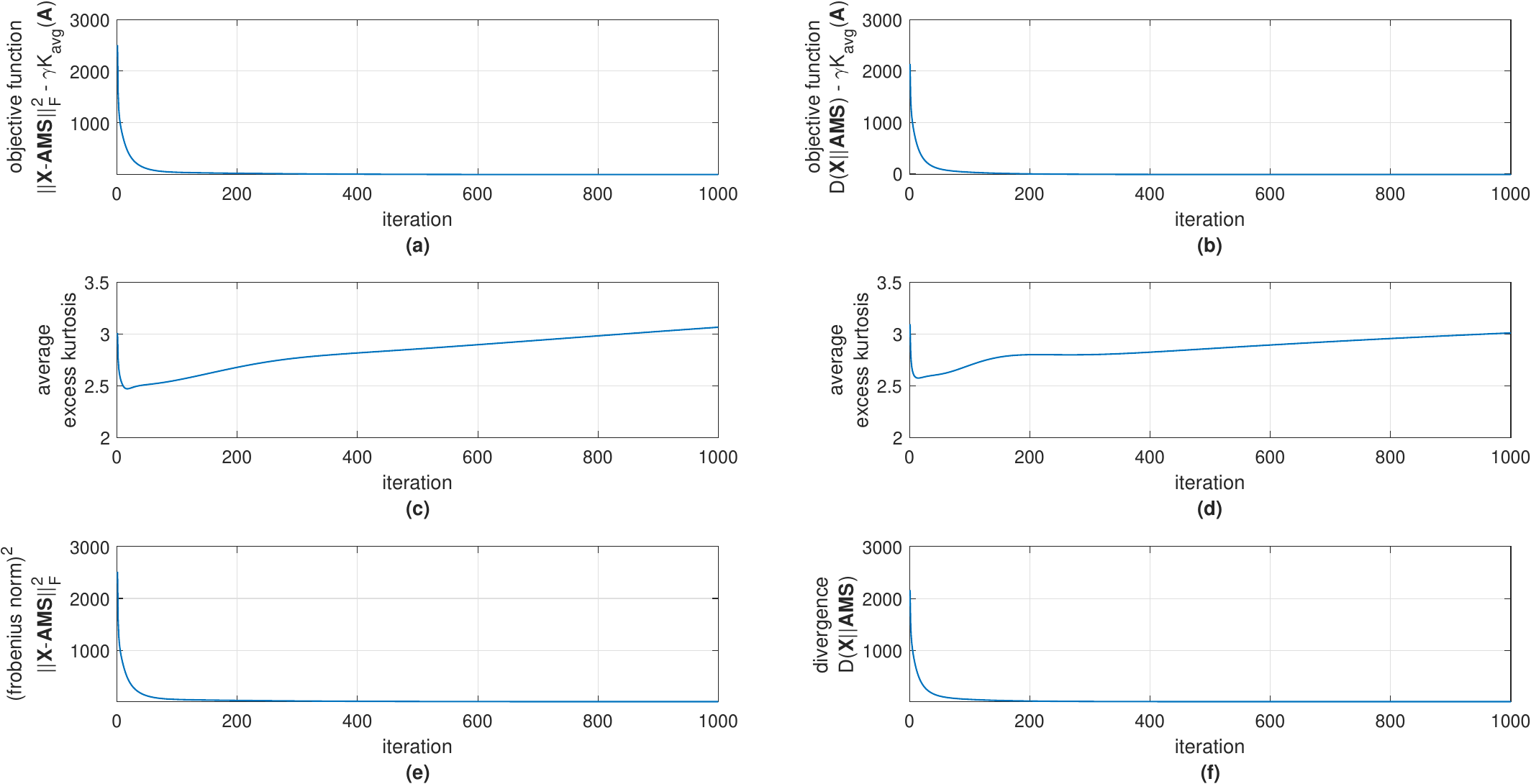}
\caption{Convergence of KbSNMF: Variation of (a) objective function, (c) average excess kurtosis of the extracted endmembers, and (e) square of Frobenius norm between X and AMS, over number iteration utilizing KbSNMFfnorm algorithm. Variation of (b) objective function, (d) average excess kurtosis of the extracted endmembers, and (f) divergence of X from AMS, over number of iteration utilizing KbSNMF-div algorithm.}
\label{convergence_KbSNMF}
\end{figure*}

In order to find a solution for (\ref{objectnew1}), we consider the objective function below.
\begin{equation} 
\label{optimization1}
\mathbb{L}(\mathbf{A},\mathbf{S}) = \Vert\mathbf{X}-\mathbf{AMS}\Vert^{2}_{F} - \gamma \overline{K}(\mathbf{A})
\end{equation}
In order to make the algorithm much simpler, the variable matrices $\mathbf{A}$ and $\mathbf{S}$ are updated in turns. In each iteration, first $\mathbf{A}$ is updated while $\mathbf{S}$ is kept constant, then, $\mathbf{S}$ is updated while $\mathbf{A}$ is kept constant. This scheme is called a block-coordinate descent approach and is widely utilized in NMF-based algorithms \cite{Burred2017DetailedDO}. The updates rules can be primarily written as follows.
\begin{equation} 
\begin{split}
\label{updateAS1}
\mathbf{A} &\leftarrow  \mathbf{A} - \mathbf{\eta_{A}} \circ \nabla_{\mathbf{A}} \mathbb{L} \\
\mathbf{S} &\leftarrow  \mathbf{S} - \mathbf{\eta_{S}} \circ \nabla_{\mathbf{S}} \mathbb{L}
\end{split}
\end{equation}
where $\circ$ denotes the Hadamard (element-by-element) product. Updating $\mathbf{A}$ and $\mathbf{S}$ directly accounts to computing the partial derivatives $\nabla_{\mathbf{A}}\mathbb{L}\in\mathbb{R}_{+}^{n\times r}$ and $\nabla_{\mathbf{S}}\mathbb{L}\in\mathbb{R}_{+}^{r\times m}$, and finding suitable learning rates $\mathbf{\eta_{A}}\in\mathbb{R}_{+}^{n\times r}$ and $\mathbf{\eta_{S}}\in\mathbb{R}_{+}^{r\times m}$. 

Computing the partial derivatives of $\mathbb{L}$ w.r.t. $\mathbf{A}$ and $\mathbf{S}$ can be seen as two parts, \textit{i.e.} partial derivatives of \mbox{$\Vert\mathbf{X}-\mathbf{AMS}\Vert^{2}_{F}$} term and $\gamma \overline{K}(\mathbf{A})$ term. We refer the readers to \cite{1580485} and \cite{Burred2017DetailedDO} for detailed explanation of the partial derivative of $\Vert\mathbf{X}-\mathbf{AMS}\Vert^{2}_{F}$. Incorporating the result in (\ref{derivation9}), we can present the partial derivatives of $\mathbb{L}$ as follows.
\begin{equation} 
\begin{split}
\label{diff}
\frac{\partial \mathbb{L}}{\partial \mathbf{A}} &= -2\mathbf{X}\mathbf{S}^{T}\mathbf{M}^{T} + 2\mathbf{A}\mathbf{M}\mathbf{S}\mathbf{S}^{T}\mathbf{M}^{T} + 2\gamma'\mathbf{N}[\mathbf{N}\mathbf{A}]^{\circ 3} \\
\frac{\partial \mathbb{L}}{\partial \mathbf{S}} &= -2\mathbf{M}^{T}\mathbf{A}^{T}\mathbf{X} + 2\mathbf{M}^{T}\mathbf{A}^{T}\mathbf{A}\mathbf{M}\mathbf{S} 
\end{split}
\end{equation}
where $\gamma'$ is the scalar quantity which equals $\frac{-2\gamma}{nr}$. By substituting $\frac{\partial \mathbb{L}}{\partial \mathbf{A}}$ and $\frac{\partial \mathbb{L}}{\partial \mathbf{S}}$ in the original block-coordinate descent equations in (\ref{updateAS1}), we can obtain the following update rules for KbSNMF-fnorm.
\small
\begin{equation} 
\begin{split}
\label{updateAS2}
\mathbf{A} &\leftarrow  \mathbf{A} - \mathbf{\eta_{A}} \circ ( -2\mathbf{X}\mathbf{S}^{T}\mathbf{M}^{T} + 2\mathbf{A}\mathbf{M}\mathbf{S}\mathbf{S}^{T}\mathbf{M}^{T} + 2\gamma'\mathbf{N}[\mathbf{N}\mathbf{A}]^{\circ 3})\\
\mathbf{S} &\leftarrow  \mathbf{S} - \mathbf{\eta_{S}} \circ (-2\mathbf{M}^{T}\mathbf{A}^{T}\mathbf{X} + 2\mathbf{M}^{T}\mathbf{A}^{T}\mathbf{A}\mathbf{M}\mathbf{S} )
\end{split}
\end{equation}
\normalsize

However, due to the subtracting terms in the gradients, the update rules (\ref{updateAS2}) can enforce $\mathbf{A}$ and $\mathbf{S}$ to contain negative elements, which contradicts with the parts-based representation of the NMF framework as well as the HU setting. Thus, following a methodology similar to that proposed by Lee and Seung \cite{2ef7006f34ff4cd7afa86c9bc8932c80}, we define data-adaptive learning rates $\mathbf{\eta_{A}}$ and $\mathbf{\eta_{S}}$ as below in order to ensure all positive elements in $\mathbf{A}$ and $\mathbf{S}$ at each update step. 
\begin{equation}
\begin{split} 
\label{learningAS}
\mathbf{\eta_{A}} &= \frac{\mathbf{A}}{2\mathbf{A}\mathbf{M}\mathbf{S}\mathbf{S}^{T}\mathbf{M}^{T} + 2\gamma'\mathbf{N}[\mathbf{N}\mathbf{A}]^{\circ 3}}\\
\mathbf{\eta_{S}} &= \frac{\mathbf{S}}{ 2\mathbf{M}^{T}\mathbf{A}^{T}\mathbf{A}\mathbf{M}\mathbf{S}}
\end{split}
\end{equation} 
The fraction line denotes element-by-element division. This results in the multiplicative update rules for the proposed KbSNMF-fnorm algorithm as below.
\begin{equation} 
\begin{split}
\label{updateAS3}
\mathbf{A}&\leftarrow\mathbf{A}\circ\frac{\mathbf{X}\mathbf{S}^{T}\mathbf{M}^{T}}{\mathbf{A}\mathbf{M}\mathbf{S}\mathbf{S}^{T}\mathbf{M}^{T} + \gamma'\mathbf{N}[\mathbf{N}\mathbf{A}]^{\circ 3} }\\
\mathbf{S}&\leftarrow \mathbf{S}\circ\frac{\mathbf{M}^{T}\mathbf{A}^{T}\mathbf{X}}{\mathbf{M}^{T}\mathbf{A}^{T}\mathbf{A}\mathbf{M}\mathbf{S}}
\end{split}
\end{equation} 
For convenience, we reconfigure the placement of matrices. Therefore, the final update rules for the proposed KbSNMF-fnorm algorithm will be as follows.
\begin{equation} 
\begin{split}
\label{updateAS4}
\mathbf{A}&\leftarrow\mathbf{A}\circ\frac{\mathbf{X(MS)}^{T}}{\mathbf{A(MS)(MS)}^{T} + \gamma'\mathbf{N}[\mathbf{N}\mathbf{A}]^{\circ 3} }\\
\mathbf{S}&\leftarrow \mathbf{S}\circ\frac{\mathbf{(AM)}^{T}\mathbf{X}}{\mathbf{(AM)}^{T}\mathbf{(AM)S}}
\end{split}
\end{equation}

It can be seen that choosing the data-adaptive learning rates in the form of (\ref{learningAS}) to avoid subtraction has enforced $\mathbf{A}$ and $\mathbf{S}$ to contain nonnegative elements throughout the block-coordinate descent approach, given initial nonnegative $\mathbf{A}$ and $\mathbf{S}$.

\subsection{KbSNMF-div} 
\label{section:KbSNMF-div}

Analogously, we present the following optimization problem for KbSNMF based on divergence (KbSNMF-div).
\begin{equation} 
\label{objectnew2}
\argmin_{\mathbf{A},\mathbf{S}}\left\{D(\mathbf{X}\Vert\mathbf{AMS}) - \gamma \overline{K}(\mathbf{A})\right\}, \hspace{10pt} s.t.\mathbf{A},\mathbf{S}\succeq 0
\end{equation}
Following a similar procedure as in Section \ref{section:KbSNMF-fnorm}, the following multiplicative update rules can be derived for KbSNMF-div algorithm,
\begin{equation} 
\begin{split}
\label{updatedivAS}
\mathbf{A}&\leftarrow\mathbf{A}\circ\frac{\frac{\mathbf{X}}{\mathbf{A(MS)}}\mathbf{(MS)}^{T}}{\mathbf{1}_{n\times m}\mathbf{(MS)}^{T} + \gamma'\mathbf{N}[\mathbf{N}\mathbf{A}]^{\circ 3}}\\
\mathbf{S}&\leftarrow\mathbf{S}\circ\frac{\mathbf{(AM)}^{T}{\frac{\mathbf{X}}{\mathbf{(AM)S}}}}{\mathbf{(AM)}^{T}\mathbf{1}_{n\times m}}
\end{split}
\end{equation} 
where $\mathbf{1}\in\mathbb{R}^{n\times m}$ is a matrix whose all elements are one, and the other notations are the same as defined previously.

\section{Algorithm Implementation} 
\label{section:Algorithm Implementation}

In this section, we will discuss several points related to the implementation of the proposed algorithm.
\begin{figure}[!t]
\center
\includegraphics[width=\columnwidth]{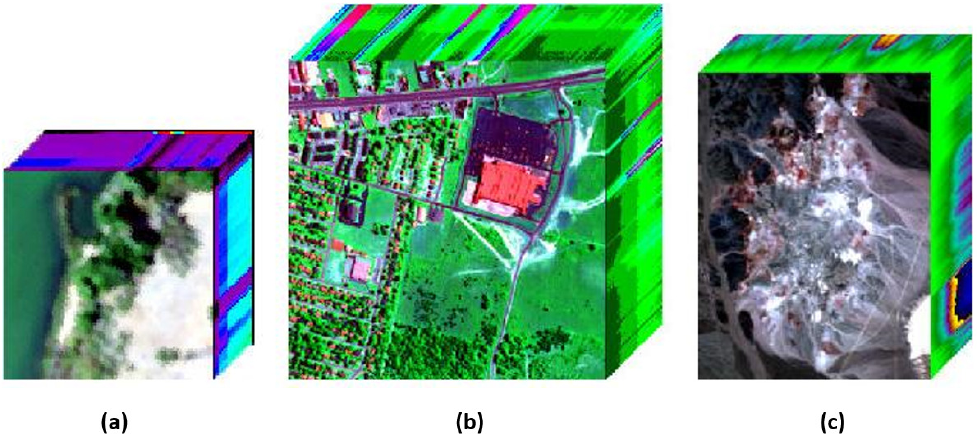}
\caption{Standard real HSI datasets: (a) Samson dataset, (b) Urban dataset, (c) Cuprite dataset.}
\label{real_datasets}
\end{figure}
\subsection{Initialization} 
\label{section:Initialization}

Many algorithms had been designed in the past to enhance the initialization of the conventional NMF problem. In this work, we utilize the Nonnegative Double Singular Value Decomposition (NNDSVD) algorithm \cite{BOUTSIDIS20081350} in order to initialize the matrices $\mathbf{A}$ and $\mathbf{S}$. NNDSVD takes the HSI $\mathbf{X}$ and the no. of endmembers $r$ as the input and generates a pair of $\mathbf{A}$ and $\mathbf{S}$ matrices. The basic NNDSVD algorithm is based on two singular value decomposition (SVD) processes, first, approximating the data matrix and the second, approximating positive sections of the resulting partial SVD factors incorporating the properties of unit rank matrices. Extensive evidence can be found to suggest that NNDSVD promotes the rapid convergence of the NMF algorithm.

\subsection{Normalization} 
\label{section:Normalization}

To avoid the complexity of computing $\nabla_{\mathbf{A}}\overline{K}$, the endmember spectra are considered as signals of unit variance (See Section \ref{section:Derivative of Average Kurtosis}), which is not always true in HU setting. In order to rectify this premise, at the beginning of each iteration of the proposed algorithm, we normalize the endmember spectra by their individual variances (See Algorithm \ref{Algorithm_main}: lines \ref{normalize line 1} and \ref{normalize line 2}). Thus, the resulting algorithm follows the essence of projected gradient descent methods which are often utilized in signal processing applications \cite{10.5555/983149}.

\subsection{Convergence}
\label{section:Convergence}

Fig. \ref{convergence_KbSNMF} demonstrates the convergence of KbSNMF over number of iterations. Here, we have fixed the parameters $\gamma$ and $\theta$ at 3 and 0.4 respectively for KbSNMF-fnorm, and at 8 and 0.4 respectively for KbSNMF-div. Selection of suitable $\gamma$ and $\theta$ and their effects on the unmixing performance are extensively discussed in Section \ref{section:Sensitivity to control parameters}. Observing Fig. \ref{convergence_KbSNMF}(a) and \ref{convergence_KbSNMF}(b), it is evident that KbSNMF convergences to a local minimum w.r.t. $\mathbf{A}$ and $\mathbf{S}$. Also our primary objective of maximizing $\overline{K}$ has been achieved, and can be clearly seen in the Fig. \ref{convergence_KbSNMF}(c) and \ref{convergence_KbSNMF}(d). In the meantime, as seen in Fig. \ref{convergence_KbSNMF}(e) and \ref{convergence_KbSNMF}(f), Frobenius norm and divergence respectively converges to local minima w.r.t. $\mathbf{A}$ and $\mathbf{S}$ which ensures the quality of approximation between $\mathbf{X}$ and $\mathbf{AMS}$.

\subsection{Termination} 
\label{section:Termination}

In this work, we utilize two stopping criteria, one based on the maximum no. of iterations and the other based on the rate of change in the objective function. We choose a maximum no. of iterations, $t_{max}$ and a minimum rate of change in the objective function $C_{min}$. The algorithm is terminated either if the present iteration $t$ reaches $t_{max}$ or if the present rate of change in the objective function $C(t)$ falls below  $C_{min}$. Here $C(t) = \frac{\mid L(t-1)-L(t)\mid}{\mid L(t-1)\mid}$, where L(t) is the value of the objective function at the $t^{\text{th}}$ iteration. The selection of suitable $t_{max}$ and $C_{min}$ is discussed in Section \ref{section:Parameter Selection}.

\subsection{Parameter Selection} 
\label{section:Parameter Selection}

Observing Fig. \ref{convergence_KbSNMF}(a) and \ref{convergence_KbSNMF}(b), it is evident that both variants of KbSNMF algorithm have converged to local minima by the $1000^{\text{th}}$ iteration. Thus, we fix $t_{max}$ at $1000$ preserving a reasonable allowance. Also it is seen that the percentage change in the objective function around the $1000^{\text{th}}$ iteration is in the order of $10^{-4}$. Thus, we fix $C_{min}$ at $10^{-5}$ to ensure convergence. Determining optimum control parameters $\gamma$ and $\theta$ is discussed in Section \ref{section:Sensitivity to control parameters} via experiment. 

Adhering to all the implementing issues discussed above, the proposed KbSNMF algorithm can be summarized as in Algorithm \ref{Algorithm_main}. 

\begin{algorithm}
\caption{KbSNMF Algorithm for HU}	
\label{Algorithm_main}
\SetAlgoLined
\KwIn{$\mathbf{X}$, $r$, $\gamma$, $\theta$, $t_{max}$, and $C_{min}$, $Algorithm\text{ }variant$}
Initialize $\mathbf{A}$ and $\mathbf{S}$ utilizing NNDSVD algorithm \cite{BOUTSIDIS20081350}\;
Compute $\mathbf{M}$ utilizing (\ref{smoothing_matrix})\;
Compute $\mathbf{N}$ as in (\ref{derivation7})\;
Compute $\gamma'$ as in (\ref{diff})\;
Normalize each column of $\mathbf{A}$ w.r.t its variance\; \label{normalize line 1}
\While{$t\leq t_{max} \land C\geq C_{min}$}{
Update $\mathbf{A}$:\\
\Switch{Algorithm variant}{
\Case{KbSNMF-fnorm}{
	$\mathbf{A}\leftarrow\mathbf{A}\circ\frac{\mathbf{X(MS)}^{T}}{\mathbf{A(MS)(MS)}^{T} + \gamma'\mathbf{N}[\mathbf{N}\mathbf{A}]^{\circ 3}}$;
}
\Case{KbSNMF-div}{
	$\mathbf{A}\leftarrow\mathbf{A}\circ\frac{\frac{\mathbf{X}}{\mathbf{A(MS)}}\mathbf{(MS)}^{T}}{\mathbf{1}_{n\times m}\mathbf{(MS)}^{T} + \gamma'\mathbf{N}[\mathbf{N}\mathbf{A}]^{\circ 3}}$;
}
}
Normalize each column of $\mathbf{A}$ w.r.t its variance\; \label{normalize line 2}
Update $\mathbf{S}$:\\
\Switch{Algorithm variant}{
\Case{KbSNMF-fnorm}{
	$\mathbf{S}\leftarrow \mathbf{S}\circ\frac{\mathbf{(AM)}^{T}\mathbf{X}}{\mathbf{(AM)}^{T}\mathbf{(AM)S}}$
}
\Case{KbSNMF-div}{
	$\mathbf{S}\leftarrow \mathbf{S}\circ\frac{\mathbf{(AM)}^{T}{\frac{\mathbf{X}}{\mathbf{(AM)S}}}}{\mathbf{(AM)}^{T}\mathbf{1}_{n\times   m}}$
}
}
}
\KwOut{Extracted $\mathbf{A}$ and $\mathbf{S}$}

\end{algorithm}
\begin{figure}[!b]
\centering
\includegraphics[width=\columnwidth]{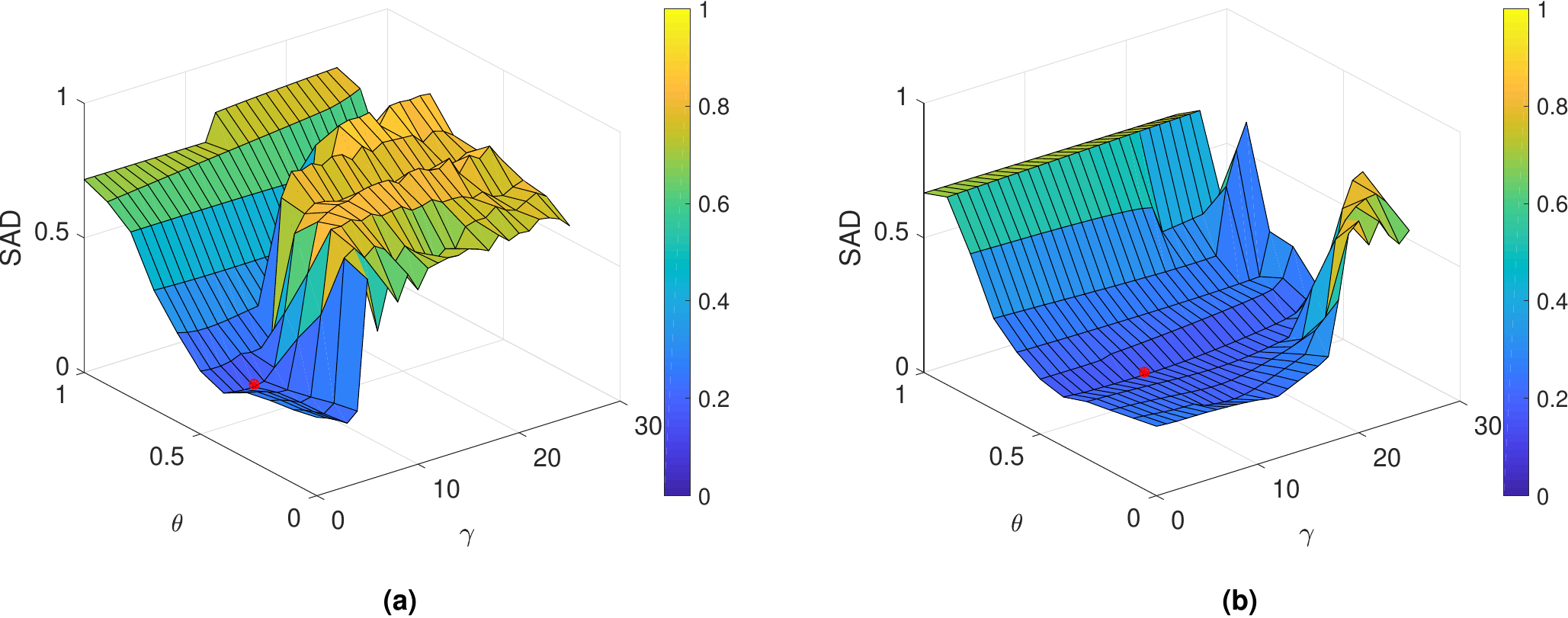}
\caption{Variation of unmixing performance in terms of SAD with $ \gamma $ and $\theta$ for (a) KbSNMF-fnorm and (b) KbSNMF-div. The minimum SAD value in each 3-D surface is marked in red.}
\label{control_parameters}
\end{figure}
\section{Experiments and Discussions} 
\label{section:Experiments and Discussions}

\subsection{Performance Criteria} 
\label{section:Performance Criteria}

In order to evaluate the performance of the proposed KbSNMF algorithm and assess its competitiveness with the other state-of-the-art algorithms, we utilize two performance criteria, which are commonly adopted in HU performance evaluation, \textit{i.e.} Spectral Angle Distance (SAD) and Root Mean Square Error (RMSE). In most of the previous literature on HU, SAD had been utilized to compare the extracted endmember spectra with the ground truth endmember spectra while RMSE had been utilized to compare the extracted abundance maps with the ground truth abundance maps. In our work $SAD_{i}$, as in (\ref{SAD}) measures the spectral angle between the $i^{\text{th}}$ ground truth endmember spectrum $\mathbf{a_{i}}$ and the corresponding extracted endmember spectrum $\mathbf{\widehat{a}_{i}}$, in radians; $RMSE_{i}$, as in (\ref{RMSE}) measures the error between the $i^{\text{th}}$ ground truth abundance map $\mathbf{S_{i}}$ and the corresponding extracted abundance map $\mathbf{\widehat{S}_{i}}$.

\begin{equation} 
\label{SAD}
SAD_{i}=\cos^{-1}\left(\frac{\widehat{\mathbf{a}}_{i}^{T}\mathbf{a}_{i}}{\Vert\widehat{\mathbf{a}}_{i}\Vert_{2}\Vert\mathbf{a}_{i}\Vert_{2}}\right)
\end{equation} 
\begin{equation} 
\label{RMSE}
RMSE_{i}=\sqrt{\frac{1}{m}\sum_{j=1}^{m}\big(\mathbf{S}_{ij}-\widehat{\mathbf{S}}_{ij}\big)^{2}}
\end{equation} 

Unless otherwise noted, in all experiments SAD and RMSE are average values over all extracted endmember spectra and abundance maps respectively.

\subsection{Experimental Setting} 
\label{section:Experimental Setting}
The proposed algorithm is tested on simulated as well as real hyperspectral datasets. Also, we compare the performance of our proposed algorithm with the popular state-of-the-art NMF-based HU baselines: $l_{1/2}$-NMF \cite{5871318}, SGSNMF \cite{7995123}, Min-vol NMF \cite{8682280}, R-CoNMF \cite{7505959}, SSRNMF \cite{9146211} and MVNTF\cite{7784711}. To ensure that the evaluations are done on common grounds, we utilize the same initializing procedure and stopping criteria as mentioned in Sections \ref{section:Initialization} and \ref{section:Termination} respectively, for all the competing algorithms except MVNTF algorithm which is initialized with random matrices. 

Simulated HSI data were generated utilizing the hyperspectral imagery synthesis toolbox (HSIST)\footnote{\url{http://www.ehu.eus/ccwintco/index.php/Hyperspectral_Imagery_Synthesis_tools_for_MATLAB}} in order to conduct experiments. HSIST consists of the full USGS spectral library\footnote{\url{https://www.usgs.gov/labs/spec-lab}} which contains hundreds of endmember spectra including minerals, organic and volatile compounds, vegetation, and man-made materials. The corresponding abundance maps were generated incorporating a spherical Gaussian field \cite{7423723}.
\begin{figure}[!b]
\centering
\includegraphics[width=\columnwidth]{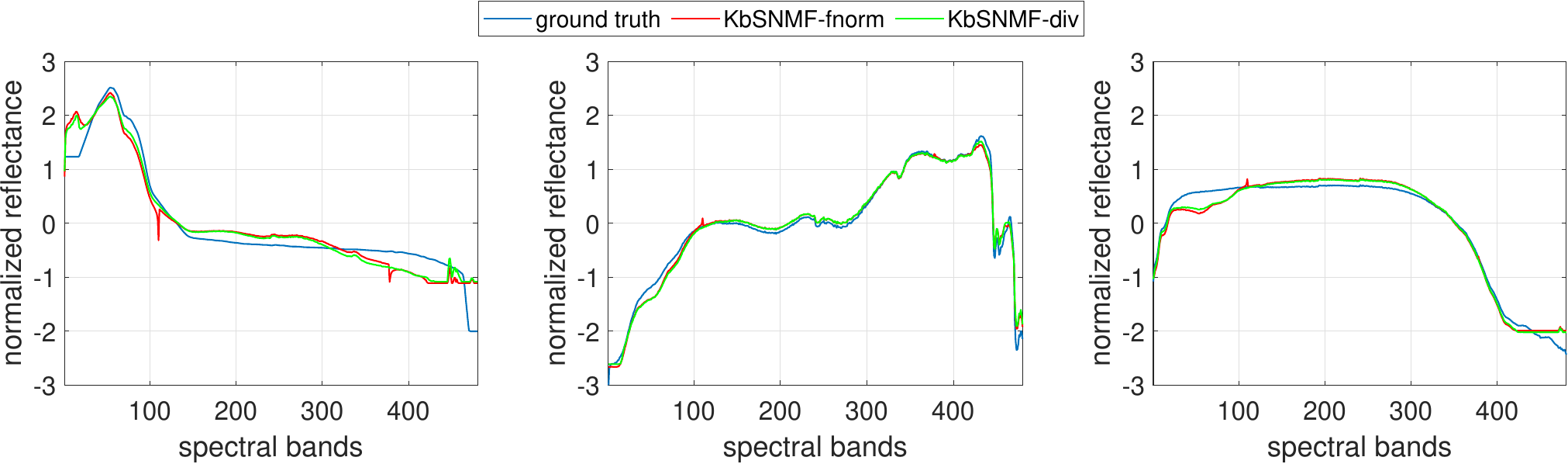}
\caption{Endmember spectra extracted utilizing KbSNMF: “Seawater”, “Clintonite” and “Sodiumbicarbonate” respectively.}
\label{performance_simulated_endmember}
\end{figure}
\begin{figure}[!b]
\centering
\includegraphics[width=\columnwidth]{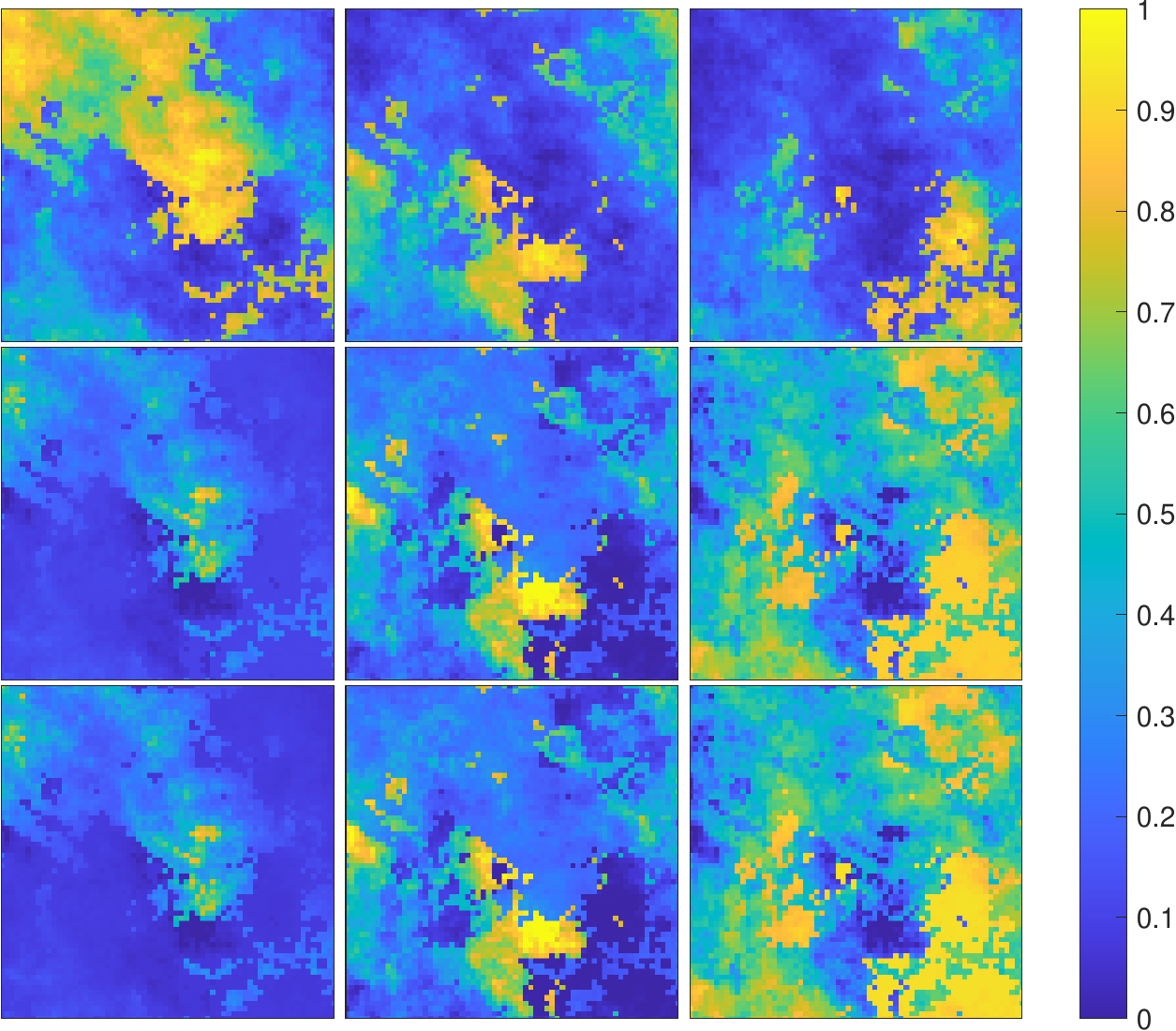}
\caption{Abundance maps extracted utilizing KbSNMF: Top row- ground truth abundance maps, Middle rowextracted
abundance maps by KbSNMF-fnorm. Bottom row- extracted abundance maps by KbSNMF-div.}
\label{performance_simulated_abundance}
\end{figure}

To assess the performance of the proposed method in real environments, we conduct experiments on real hyperspectral data. The Samson dataset, the Urban dataset, and the Cuprite dataset (See Fig. \ref{real_datasets}) have been widely utilized for performance evaluation and comparison in recent HU studies \cite{lu2020subspace, qin2020blind, 9096565}. The Samson dataset's each pixel is recorded at 156 spectral channels covering wavelengths in the range of 401-889 nm with a spectral resolution of 3.13 nm. The Urban dataset's each pixel is recorded at 210 spectral channels originally, however, due to dense water vapor and atmospheric effects, several bands are customarily removed prior to analysis, resulting in 162 spectral bands ranging from 400-2500 nm, with a spectral resolution of 10 nm. The Urban dataset possesses several ground truth versions, here we utilize the one with five endmembers. The Cuprite dataset is the widely used benchmark dataset for HU and its each pixel is recorded at 188 spectral channels covering wavelengths in the range of 370-2480 nm.

The ground truths for all real datasets are worked out utilizing a procedure similar to that of \cite{4694061} and \cite{4358857}. First, the Virtual Dimensionality (VD) algorithm \cite{1273593} is utilized to determine the no. of endmembers of the HSI. Second, the pixels that contain pure endmember spectra are chosen manually in accordance with the USGS mineral spectral library. Finally, the corresponding abundances are computed utilizing the CVX optimization Toolbox in MATLAB. Accordingly-generated ground truths are often utilized in HU method evaluation and comparison and are readily-available\footnote{\url{http://www.ehu.eus/ccwintco/index.php/Hyperspectral_Remote_Sensing_Scenes}}.

\subsection{Experiments on simulated data} 
\label{section:Experiments on simulated data}
\subsubsection{Sensitivity to control parameters}
\label{section:Sensitivity to control parameters}

We conduct experiments to find optimum values for $\gamma$ and $\theta$ for KbSNMF-fnorm and KbSNMF-div. We increase $\gamma$ from 0 to 25 in steps of 1, increase $\theta$ from 0 to 1 in steps of 0.1, and evaluate the unmixing performance at each step. It is seen that SAD reaches minimum around $\gamma = 3$ and $\theta = 0.4$ in Fig. \ref{control_parameters}(a) and around $\gamma = 8$ and $\theta = 0.4$ in Fig. \ref{control_parameters}(b). Thus, we fix $\gamma$ and $\theta$ at 3 and 0.4 respectively for KbSNMF-fnorm and at 8 and 0.4 respectively for KbSNMF-div.

\subsubsection{Unmixing performance}
\label{section:Unmixing performance}

Under this experiment, we compare the unmixing performance of KbSNMF with the other HU algorithms. Table \ref{table_performance_simulated_SAD} shows SAD values for each of the extracted endmember spectra and Table \ref{table_performance_simulated_RMSE} shows RMSE values for each of the extracted abundance maps, under the different methods. It is clearly seen that the KbSNMF under its both variants dominates the other competing algorithms in terms of SAD while signifying competitive performance in terms of RMSE. Fig. \ref{performance_simulated_endmember} and \ref{performance_simulated_abundance} respectively illustrate the endmember spectra and abundance maps extracted utilizing KbSNMF along with their ground truths.

\subsubsection{Robustness to noise}
\label{section:Robustness to noise}

\begin{figure}[!b]
\centering
\includegraphics[width=\columnwidth]{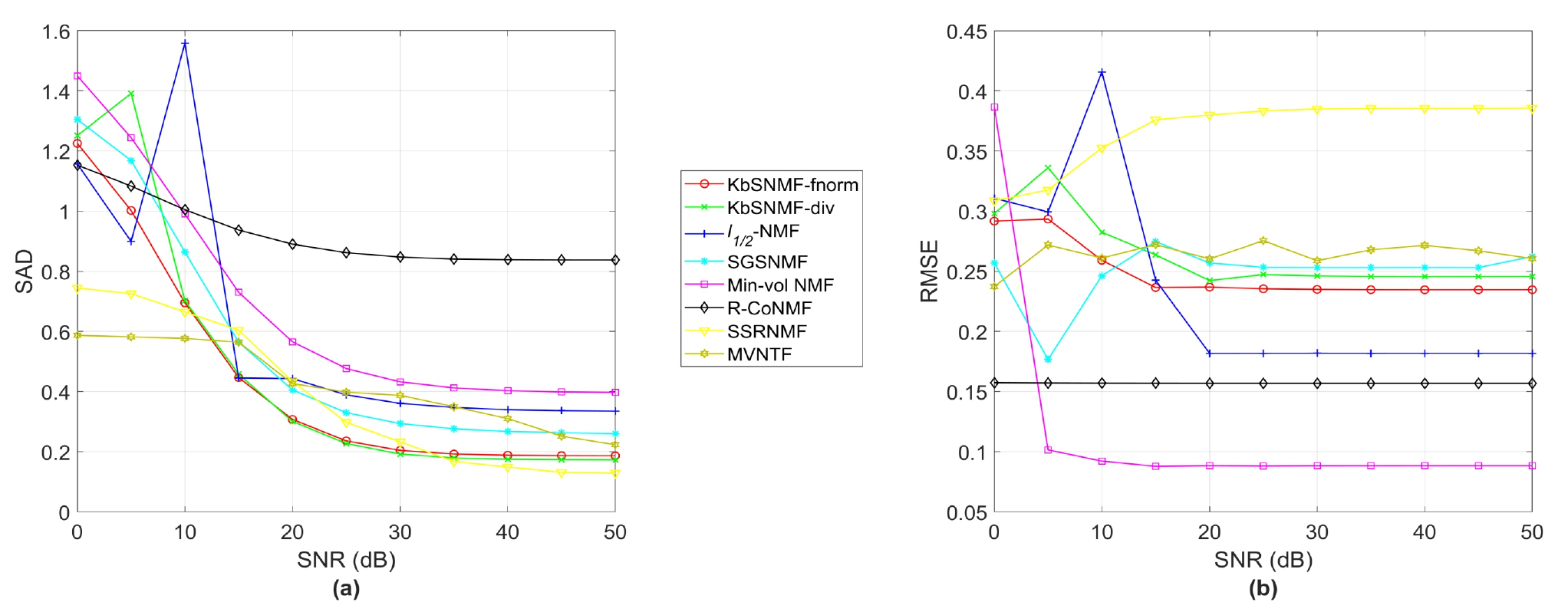}
\caption{Variation of (a) SAD and (b) RMSE with the noise level}
\label{SNR}
\end{figure}

In this experiment, we aim to analyze how the proposed algorithm performs in noisy environments. We add zero-mean white Gaussian noise to the original noise-free simulated dataset with a predetermined signal to noise ratio (SNR) given by (\ref{SNR_eq}),
\begin{equation} 
\label{SNR_eq}
\text{SNR} = 10\log_{10}\frac{\mathbb{E}(\mathbf{x}^{T}\mathbf{x})}{\mathbb{E}(\mathbf{n}^{T}\mathbf{n})} 
\end{equation} 
where $\mathbf{x}$ is the pixel spectrum vector, $\mathbf{n}$ is the noise signal vector, and $\mathbb{E}$ is the expectation operator. We conduct the experiment under 11 SNR levels: 0 dB, 5 dB, 10 dB, 15 dB, 20 dB, 25 dB, 30 dB, 35 dB, 40 dB, 45 dB, 50 dB and the results are illustrated in Fig. \ref{SNR} in terms of SAD and RMSE.Although SSRNMF and MVNTF show high immunity to large noise in terms of SAD values, It is discrenible that KbSNMF-fnorm and KbSNMF-div report the best performance showing superior performance over all competing algorithms at noise levels in the range of 15-50 dB. They also show robustness to noise up until 30 dB. In terms of RMSE, both KbSNMF-fnorm and KbSNMF-div show robustness to noise up until 20 dB and gradually deteriorate in performance thereafter. However, both KbSNMF-fnorm and KbSNMF-div outperform SSRNMF at all noise levels in terms of RMSE. The superior performance of KbSNMF-fnorm and KbSNMF-div in terms of SAD is due to the novel auxiliary regularizes on the endmember matrix and thereby attempting to extract the most realistic endmember spectra.

\begin{table*}[!b]
\centering
\captionsetup{justification=centering, labelsep=newline}
\caption{Unmixing performance comparison in terms of SAD for the simulated dataset. The best performances are in bold typeface; the second best performances are italicized; and the third best performances are underlined.}
\resizebox{\textwidth}{!}{
\begin{tabular}{c c c c c c c c c}
\hline \hline
Methods 			& \begin{tabular}{c} KbSNMF\\[-0.8ex] fnorm \end{tabular}& \begin{tabular}{c} KbSNMF\\[-0.8ex] div \end{tabular} & $l_{1/2}$-NMF & SGSNMF & \begin{tabular}{c}Min-vol\\[-0.8ex] NMF \end{tabular} & R-CoNMF  & SSRNMF & MVNTF\\
\hline
Seawater 			& \textit{0.3000} & \textbf{0.2746} & 0.6855 & 0.5416 & 0.8891 & 1.7408  & \underline{0.3017} & 0.3905\\
Clintonite 			& \textbf{0.1078} & \textit{0.1119} & 0.2990 & 0.2288 & 0.2803 & 0.7109  &0.1410& \underline{0.1326}\\
Sodiumbicarbonate 	& 0.1508 & 0.1311 & \underline{0.0168} & \textbf{0.0090} & 0.0188 & 0.0614  & \textit{0.0160} & 0.0567\\
\hline
Average             & \underline{0.1862} & \textit{0.1725} & 0.3337 & 0.2598 & 0.3961 & 0.8377  & \textbf{0.1529} & 0.1933\\
\hline \hline
\end{tabular}}
\label{table_performance_simulated_SAD}
\end{table*}


\begin{table*}[!b]
\centering
\captionsetup{justification=centering, labelsep=newline}
\caption{Unmixing performance comparison in terms of RMSE for the simulated dataset. The best performances are in bold typeface; the second best performances are italicized; and the third best performances are underlined.}
\resizebox{\textwidth}{!}{
\begin{tabular}{c c c c c c c c c}
\hline \hline
Methods 			& \begin{tabular}{c} KbSNMF\\[-0.8ex] fnorm \end{tabular}& \begin{tabular}{c} KbSNMF\\[-0.8ex] div \end{tabular} & $l_{1/2}$-NMF & SGSNMF & \begin{tabular}{c}Min-vol\\[-0.8ex] NMF \end{tabular} & R-CoNMF  & SSRNMF & MVNTF\\
\hline
Seawater 			& 0.3049 & 0.3190 & 0.2488 & 0.3579 & \textit{0.1052} & \underline{0.1753}  &\textbf{0.0817} & 0.4088\\
Clintonite 			& 0.1195 & 0.1200 & \textbf{0.0573} & 0.2906 & \underline{0.0854} & 0.1473  &\textit{0.0774} & 0.1506\\
Sodiumbicarbonate 	& 0.2797 & 0.2985 & 0.2394 & \underline{0.1123} & \textbf{0.0745} & 0.1475  &\textit{0.0768} & 0.1324\\
\hline
Average             & 0.2347 & 0.2458 & \underline{0.1818} & 0.2538 & \textit{0.0884} & 0.1569 &\textbf{0.0786} & 0.2306\\
\hline \hline
\end{tabular}}
\label{table_performance_simulated_RMSE}
\end{table*}
\begin{figure}[!b]
\centering
\includegraphics[width=\columnwidth]{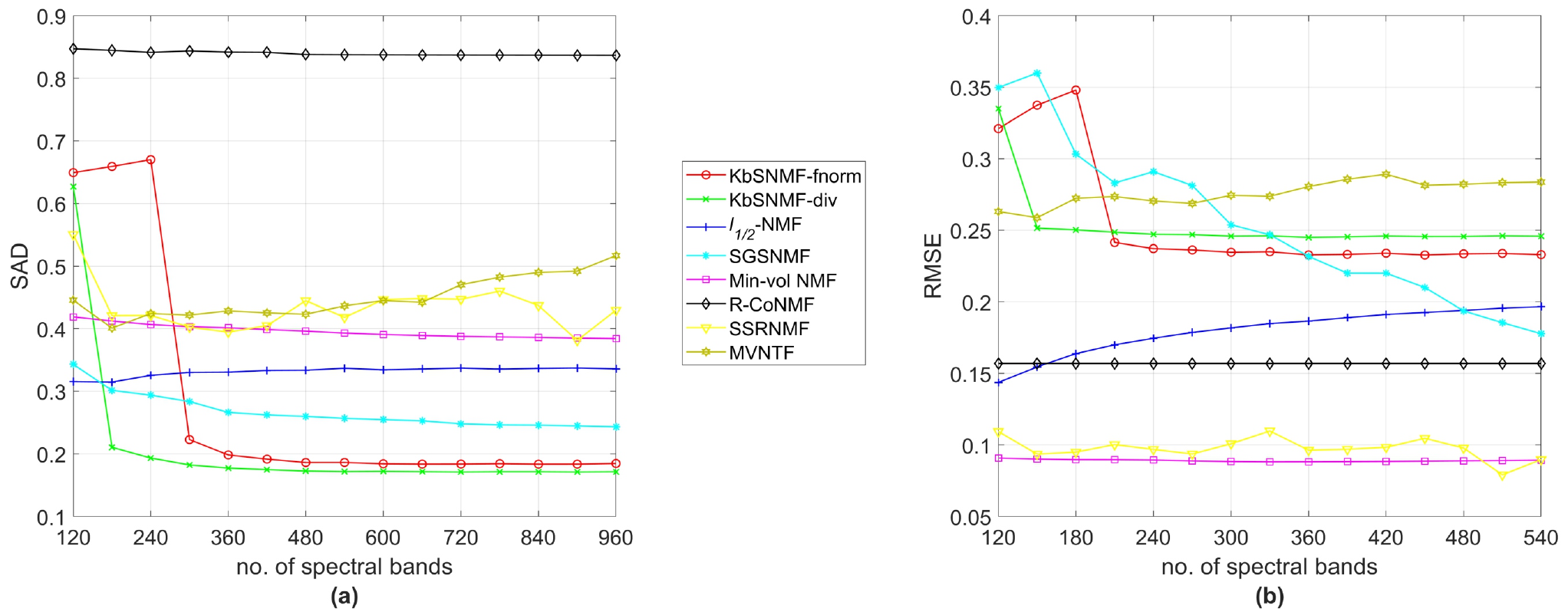}
\caption{Variation of (a) SAD and (b) RMSE with the no. of spectral bands}
\label{spectral}
\end{figure}
\begin{figure}[!b]
\centering
\includegraphics[width=\columnwidth]{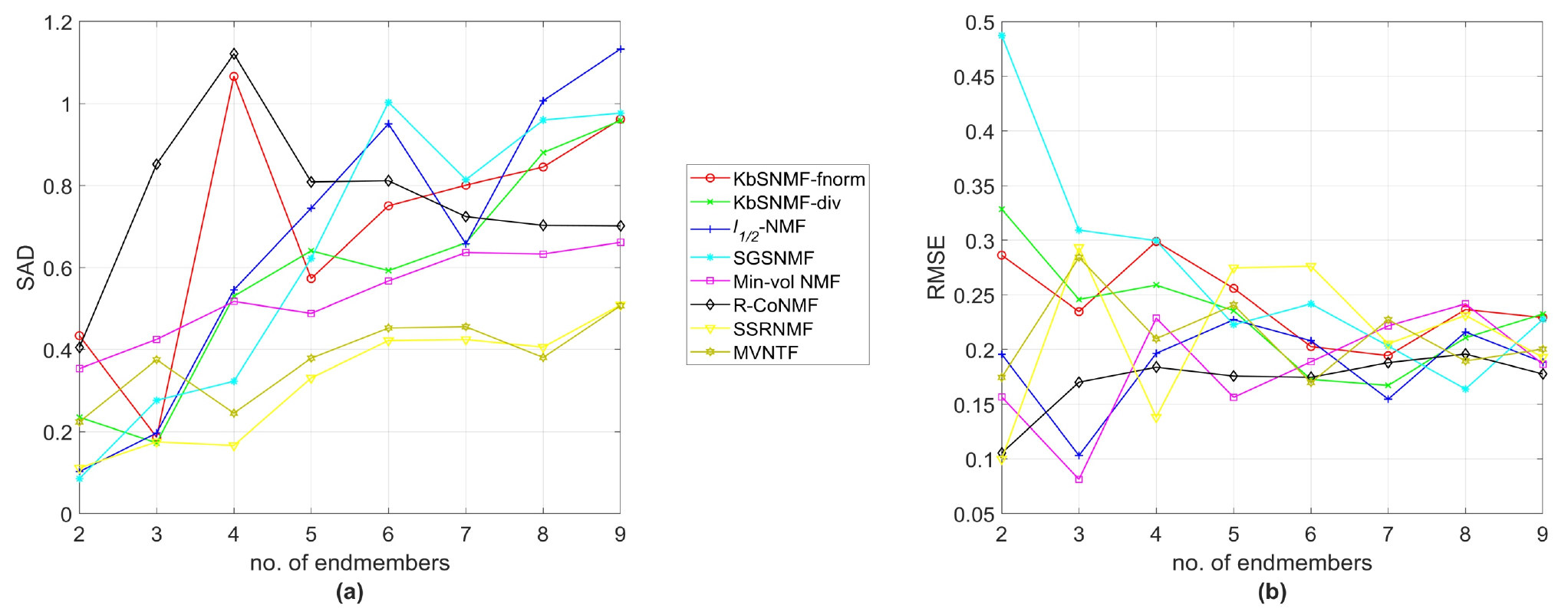}
\caption{Variation of (a) SAD and (b) RMSE with the no. of endmembers}
\label{endmembers}
\end{figure}
\begin{figure}[!b]
\centering
\includegraphics[width=\columnwidth]{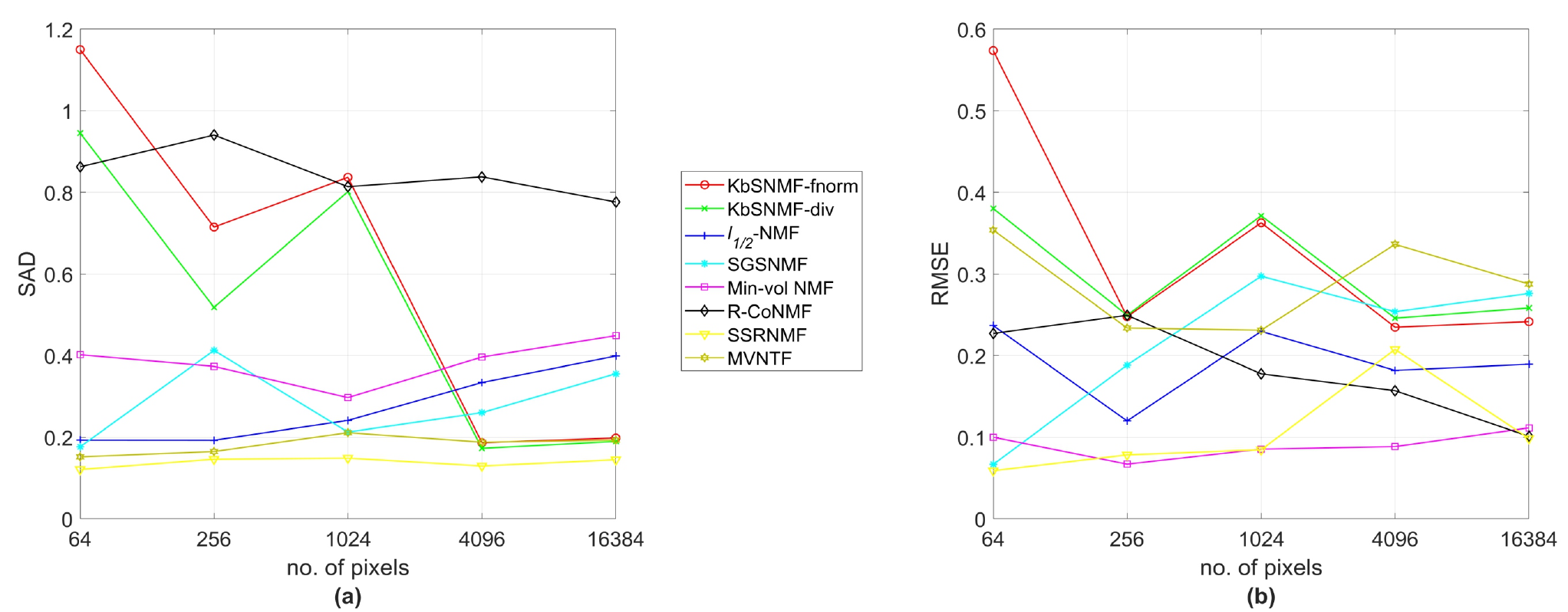}
\caption{Variation of (a) SAD and (b) RMSE with the no. of pixels}
\label{pixels}
\end{figure}

\subsubsection{Sensitivity to number of spectral bands}
\label{section:Sensitivity to number of spectral bands}

Here we vary the no. of spectral bands of the endmembers and observe the unmixing performance of the algorithms. The results are shown in Fig. \ref{spectral}. KbSNMF-fnorm and KbSNMF-div outperform all the competing algorithms in terms of SAD for no. of spectral bands in the range of 300-960. However, the performance of KbSNMF-fnorm and KbSNMF-div deteriorate drastically in terms of SAD for very low no. of spectral bands, \textit{i.e.} around 200 spectral bands. In terms of RMSE, KbSNMF-fnorm and KbSNMF-div outperforms MVNTF at high no. of spectral bands, specifically more than 180, and outperform SGSNMF at low no. of spectral bands, \textit{i.e.} below 480 spectral bands.

\begin{figure}[!t]
\centering
\includegraphics[width=\columnwidth]{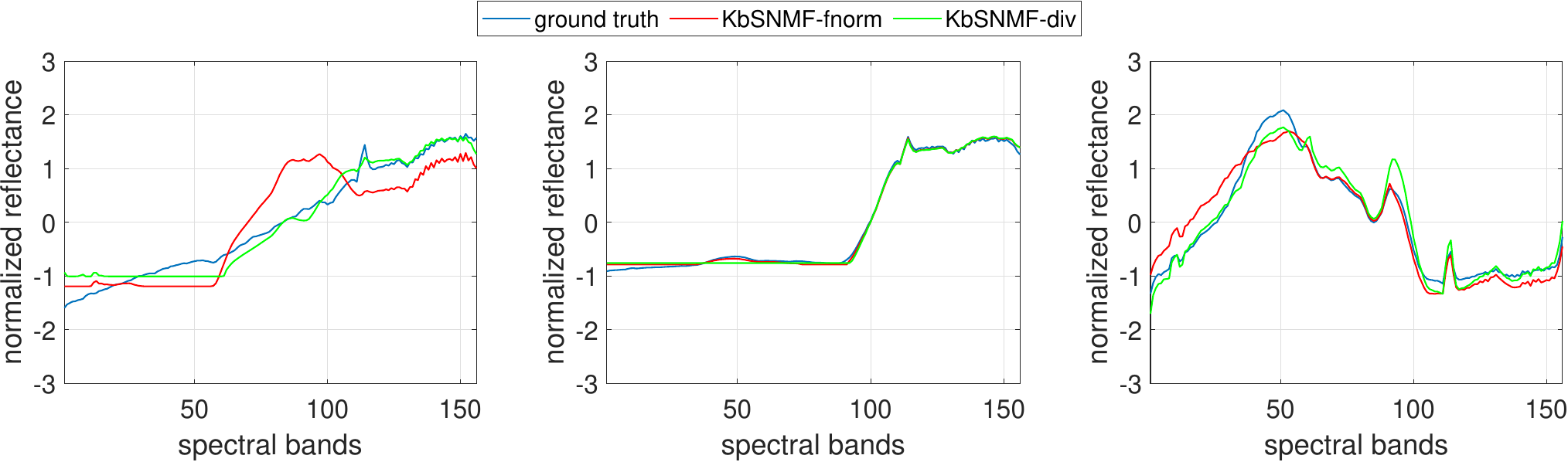}
\caption{Endmember spectra extracted utilizing Kb- SNMF of the Samson dataset: “Soil”, “Tree” and “Water” respectively}
\label{performance_samson_endmember}
\end{figure}
\begin{figure}[!t]
\centering
\includegraphics[width=\columnwidth,height=7cm]{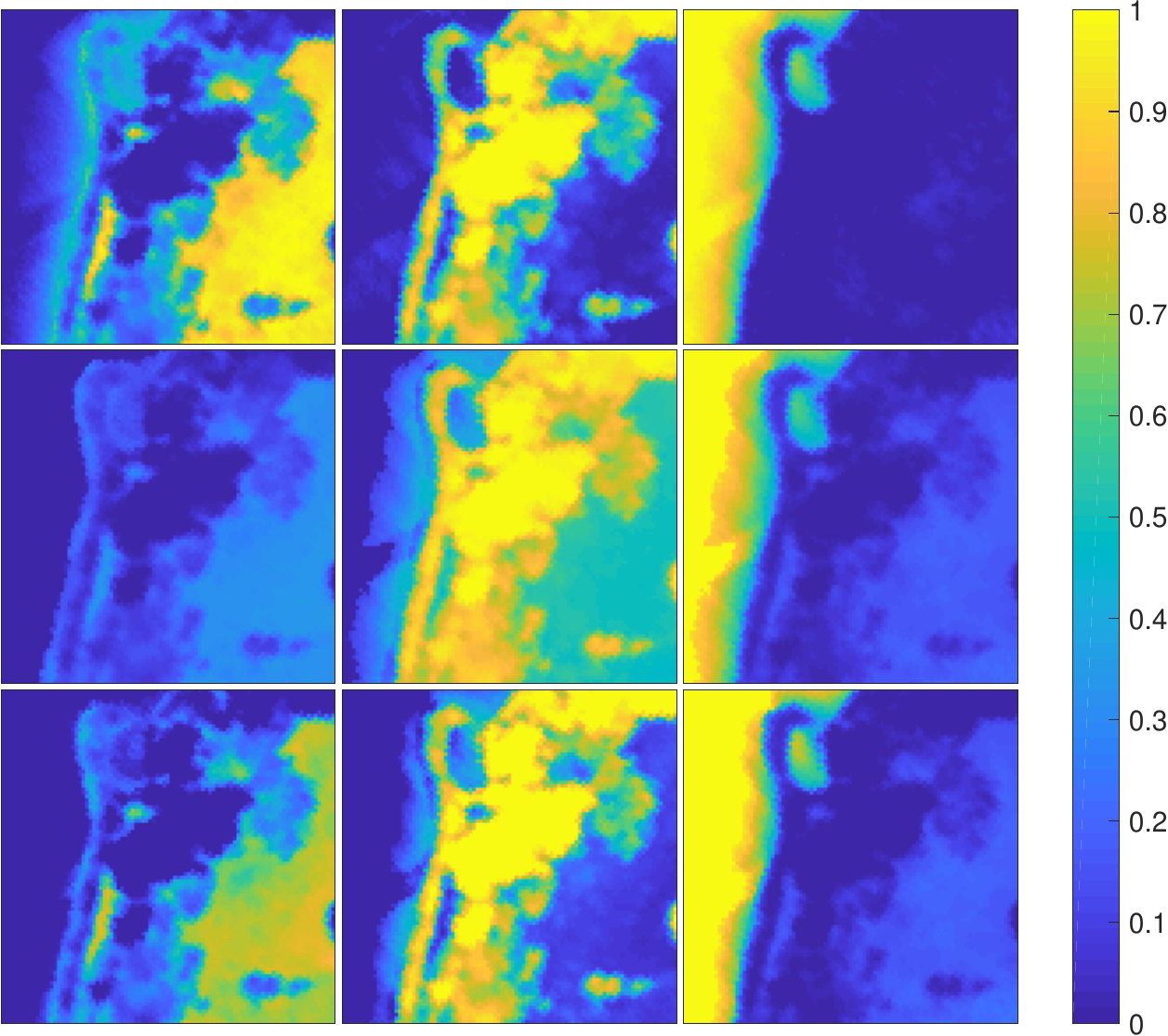}
\caption{Abundance maps extracted utilizing KbSNMF of the Samson dataset: Top row- ground truth abundance maps, Middle row- extracted abundance maps by KbSNMF-fnorm. Bottom row- extracted abundance maps by KbSNMF-div.}
\label{performance_samson_abundance}
\end{figure}

\subsubsection{Sensitivity to number of endmembers}
\label{section:Sensitivity to number of endmembers}

In this experiment, we vary the no. of endmembers and investigate the performance of the algorithms. The results are illustrated in Fig. \ref{endmembers}. All the algorithms have the tendency to deteriorate performance in terms of SAD with the no. of endmembers. KbSNMF-fnorm and KbSNMF-div outperform R-CoNMF when the no. of endmembers are low, \textit{i.e.} below 7 endmembers, and outperform SGSNMF when the no. of endmembers is high, \textit{i.e.} above 5 endmembers. In terms of RMSE, KbSNMF-fnorm and KbSNMF-div outperform SGSNMF when the no. of endmembers is low, \textit{i.e.} below 4 endmembers.

\subsubsection{Sensitivity to number of pixels}
\label{section:Sensitivity to number of pixels}

Within this experiment, we illustrate how the proposed algorithm performs under simulated HSI datasets against different no. of pixels. The no. of pixels in an HSI is a major concern since it denotes the amount of statistical information in the input to the algorithm. The amount of statistical information presented to a numerical algorithm determines the tendency of an algorithm to be trapped in a local minima \cite{7373556}. Fig. \ref{pixels} illustrates the results in terms of SAD and RMSE. The unmixing performance of KbSNMF-fnorm and KbSNMF-div improves in terms of SAD when the no. of pixels is increased and even outperforms all competing algorithms except MVNTF and SSRNMF when the no. of pixels is very high, \textit{i.e.} 64$\times$64 and 128$\times$128 pixels. In terms of RMSE, KbSNMF-fnorm and KbSNMF-div outperform SGSNMF when the no. of pixels is very high, \textit{i.e.} 64$\times$64 and 128$\times$128 pixels.

\begin{table*}[!b]
\centering
\captionsetup{justification=centering, labelsep=newline}
\caption{Unmixing performance comparison in terms of SAD for the Samson dataset. The best performances are in bold typeface; the second best performances are italicized; and the third best performances are underlined.}
\resizebox{\textwidth}{!}{
\begin{tabular}{c c c c c c c c c}
\hline \hline
Methods 			& \begin{tabular}{c} KbSNMF\\[-0.8ex] fnorm \end{tabular}& \begin{tabular}{c} KbSNMF\\[-0.8ex] div \end{tabular} & $l_{1/2}$-NMF & SGSNMF & \begin{tabular}{c}Min-vol\\[-0.8ex] NMF \end{tabular} & R-CoNMF  & SSRNMF & MVNTF\\
\hline
Soil 				& 0.4975 & \underline{0.2078} & 0.3455 & 0.3743 & 0.3463 & \textbf{0.1253}  &0.6091 & \textit{0.1488}\\
Tree 				& \textit{0.0456} & \underline{0.0647} & 0.1433  & 0.1721  & 0.2315  & \textbf{0.0105} 	&0.0750 & 0.0944\\
Water 				& 0.2771 & \underline{0.2014} & 0.3513 & 0.2941 & 0.2429 & 0.3219 &\textit{0.1624}	& \textbf{0.0887} \\
\hline
Average				& 0.2734 & \underline{0.1580} & 0.2800 & 0.2802 & 0.2736& \textit{0.1526}  &0.2822 & \textbf{0.1106}\\
\hline \hline
\end{tabular}}
\label{table_performance_samson_SAD}
\end{table*}
\begin{table*}[!b]
\centering
\captionsetup{justification=centering, labelsep=newline}
\caption{Unmixing performance comparison in terms of RMSE for the Samson dataset; The best performances are in bold typeface; the second best performances are italicized, and the third best performances are underlined.}
\resizebox{\textwidth}{!}{
\begin{tabular}{c c c c c c c c c}
\hline \hline
Methods 			& \begin{tabular}{c} KbSNMF\\[-0.8ex] fnorm \end{tabular}& \begin{tabular}{c} KbSNMF\\[-0.8ex] div \end{tabular} & $l_{1/2}$-NMF & SGSNMF & \begin{tabular}{c}Min-vol\\[-0.8ex] NMF \end{tabular} & R-CoNMF  & SSRNMF & MVNTF\\
\hline
Soil 				& 0.3429 & 0.1574 & 0.4217 & \textit{0.0532} & \underline{0.0967} & \textbf{0.0431} & 0.3832 & 0.3517	\\
Tree 				& 0.2673 & 0.0911 & \underline{0.0432} & 0.0882 & 0.1245 & \textbf{0.0118} & \textit{0.0325} &0.2454	\\
Water 				& \textit{0.0910} & \underline{0.0927} & 0.2359 & 0.1432 & \textbf{0.0432} & 0.5321 & 0.1654 & 0.4162\\
\hline
Average				& 0.2337 & \underline{0.1137} & 0.2336 & \textit{0.0949} & \textbf{0.0881} & 0.1957 & 0.1937 & 0.3378\\
\hline \hline
\end{tabular}}
\label{table_performance_samson_RMSE}
\end{table*}

\subsection{Experiments on real data} 
\label{section:Experiments on real data}

We compare the unmixing performance of KbSNMF with the other competing methods in terms of SAD and RMSE for the Samson and Urban datasets. But for the Cuprite dataset, only the SAD values are tabulated.
\begin{figure}[!t]
\centering
\includegraphics[width=\columnwidth]{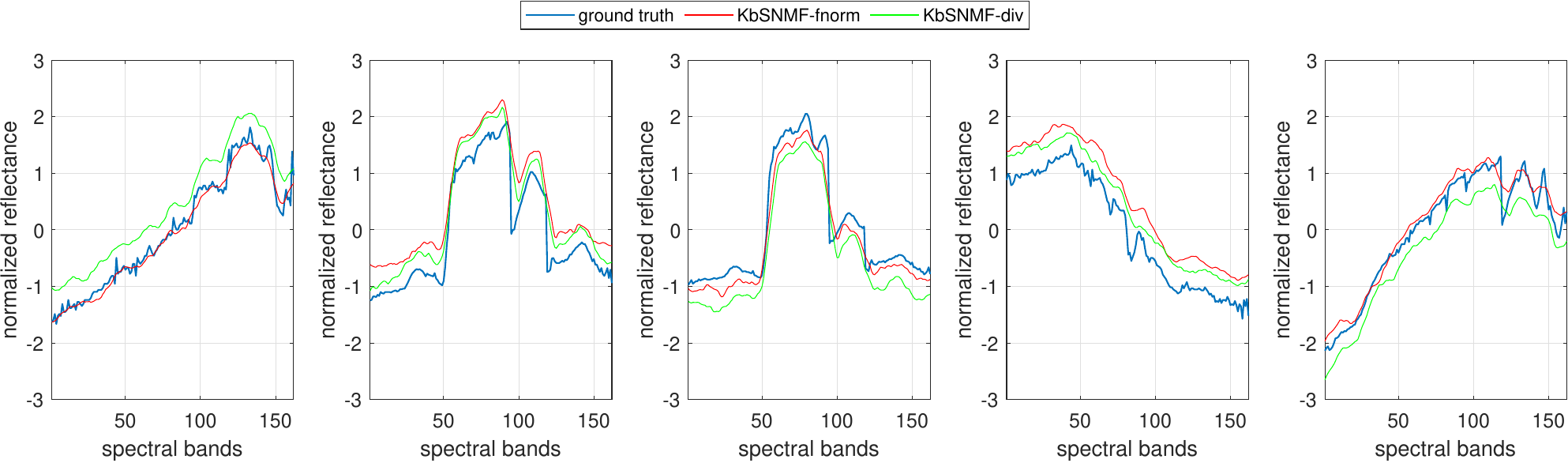}
\caption{Endmember spectra extracted utilizing Kb- SNMF of the Urban dataset: “Asphalt”, “Grass”, “Tree”, “Roof” and “Dirt” respectively.}
\label{performance_urban_endmember}
\end{figure}
\begin{figure}[!t]
\centering
\includegraphics[width=\columnwidth]{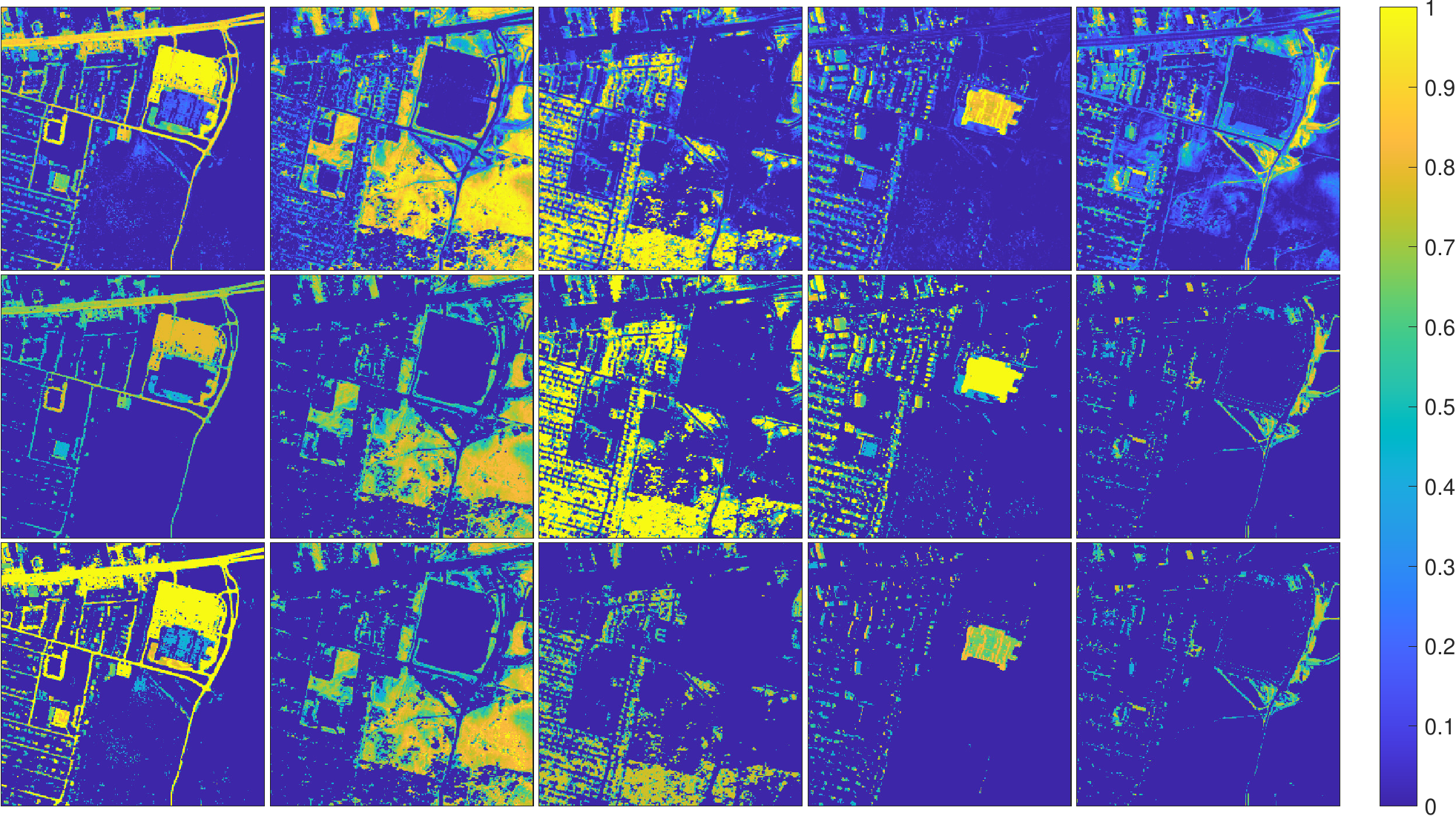}
\caption{Abundance maps extracted utilizing KbSNMF of the Urban dataset: Top row- ground truth abundance maps, Middle row- extracted abundance maps by KbSNMF-fnorm. Bottom row- extracted abundance maps by KbSNMF-div.}
\label{performance_urban_abundance}
\end{figure}

\subsubsection{Samson dataset} 
\label{section:Samson dataset}
Table \ref{table_performance_samson_SAD} shows SAD values for each of the extracted endmember spectra and Table \ref{table_performance_samson_RMSE} shows RMSE values for each of the extracted abundance maps, under the different methods. In terms of average SAD, MVNTF and R-CoNMF outperforms all methods. However, KbSNMF-fnorm and KbSNMF-div outperform the rest of the other methods. Also, KbSNMF-div reports the third best performance in terms of SAD in extracting each endmember. In terms of RMSE, Min-vol NMF and SGSNMF outperform all methods. KbSNMF-div reports the third best average performance in terms of RMSE. The endmember spectra extracted utilizing KbSNMF-fnorm and KbSNMF-div are shown in Fig. \ref{performance_samson_endmember}, and it can be observed that they closely follow their ground truth spectra. Also, the abundance maps extracted by KbSNMF-fnorm and KbSNMF-div are shown in Fig. \ref{performance_samson_abundance}, and it is evident that KbSNMF-div has managed to accurately extract the abundance maps.

\begin{figure}[!t]
\centering
\includegraphics[width=\columnwidth]{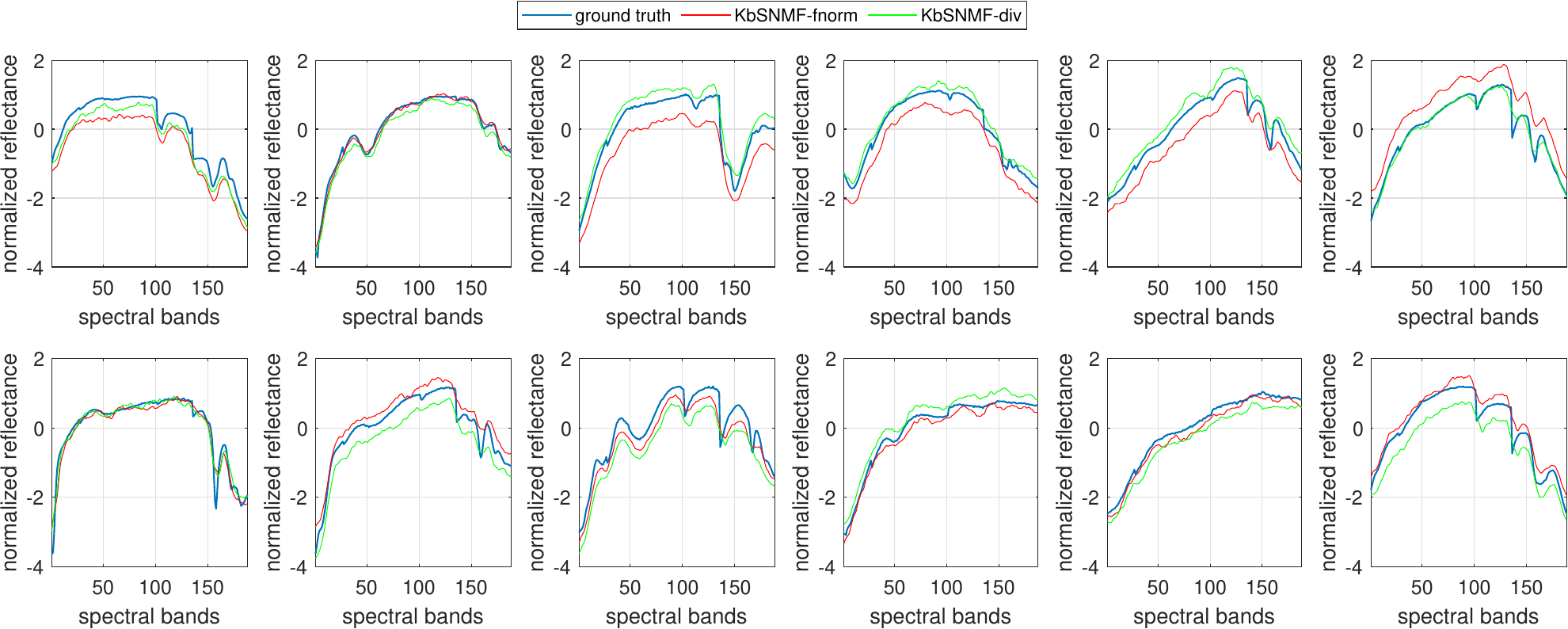}
\caption{Endmember spectra extracted utilizing Kb- SNMF of the Cuprite dataset: “Alunite”, “Andradite”,
“Buddingtonite”, “Dumortierite”, “Kaolinite 1”, “Kaolinite 2”, “Muscovite”, “Montmorillonite”, “Nontronite”, “Pyrope”, “Sphene” and “Chalcedony” respectively.}
\label{performance_cuprite_endmember}
\end{figure}
\begin{table*}[!t]
\centering
\captionsetup{justification=centering, labelsep=newline}
\caption{Unmixing performance comparison in terms of SAD for the Urban dataset. The best performances are in bold typeface; the second best performances are italicized; and the third best performances are underlined.}
\resizebox{\textwidth}{!}{
\begin{tabular}{c c c c c c c c c}
\hline \hline
Methods 			& \begin{tabular}{c} KbSNMF\\[-0.8ex] fnorm \end{tabular}& \begin{tabular}{c} KbSNMF\\[-0.8ex] div \end{tabular} & $l_{1/2}$-NMF & SGSNMF & \begin{tabular}{c}Min-vol\\[-0.8ex] NMF \end{tabular} & R-CoNMF  & SSRNMF & MVNTF\\
\hline
Asphalt 			& \textit{0.1178}& \underline{0.1252} & 0.2966 & 0.4173 & 0.1849 & 0.2017  & \textbf{0.0599} & 0.1721\\
Grass				& \textit{0.1299} & 0.2821 & 0.4993& 0.3434&\textbf{0.1045} & 0.2786  & \underline{0.1780} & 0.2080	\\
Tree 				& \textbf{0.1069}&\textit{0.1498} &0.1603 &\underline{0.1499} & 0.1798 & 0.2125 & 0.1795  & 0.2310\\
Roof 				& \textbf{0.1044}& \textit{0.1621} & 0.2518 & 0.3822& \underline{0.1930} & 0.2478  & 0.4217	& 0.3941\\
Dirt 				& 0.2999&\underline{0.1742} & 0.3379 & 0.3359&\textbf{0.1521} & 0.2435 & \textit{0.1534} & 0.3689	\\
\hline
Average				& \textbf{0.1518}& \underline{0.1787} & 0.3092 & 0.3257 & \textit{0.1628} & 0.2368  & 0.1985 & 0.2748\\
\hline \hline
\end{tabular}}
\label{table_performance_urban_SAD}
\end{table*}
\begin{table*}[!t]
\centering
\captionsetup{justification=centering, labelsep=newline}
\caption{Unmixing performance comparison in terms of RMSE for the Urban dataset; The best performances are in bold typeface; the second best performances are italicized, and the third best performances are underlined.}
\resizebox{\textwidth}{!}{
\begin{tabular}{c c c c c c c c c}
\hline \hline
Methods 			& \begin{tabular}{c} KbSNMF\\[-0.8ex] fnorm \end{tabular}& \begin{tabular}{c} KbSNMF\\[-0.8ex] div \end{tabular} & $l_{1/2}$-NMF & SGSNMF & \begin{tabular}{c}Min-vol\\[-0.8ex] NMF \end{tabular} & R-CoNMF  & SSRNMF & MVNTF\\
\hline
Asphalt 			& 0.5425 & \textit{0.1654} & 0.3197 & 0.5578 & \textbf{0.1376} & \underline{0.1980}  & 0.3237 & 0.3354\\
Grass				& \textbf{0.1296} & 0.2354 & \underline{0.2140} & 0.2545 & \textit{0.1490} & 0.3010  & 0.2437 & 0.3164	\\
Tree 				& 0.5470 & 0.4240 & 0.4607 & 0.2692 & \textbf{0.1064}& \textit{0.1258}   & \underline{0.2660} & 0.3183\\
Roof 				& \textbf{0.1528}& 0.4584 & 0.5350 & 0.2256 & 0.2130 & \underline{0.1986}  & 0.2851 & \textit{0.1858}	\\
Dirt 				& \underline{0.3691} & 0.5789 & 0.4293 & 0.5610 & 0.5409 & 0.5210  &\textit{0.3589} & \textbf{0.2004}	\\
\hline
Average				& 0.3482 & 0.3724 & 0.3917 & 0.3736 & \textbf{0.2294} & \textit{0.2689}  & 0.3121 & \underline{0.2713}	\\
\hline \hline
\end{tabular}}
\label{table_performance_urban_RMSE}
\end{table*}

\begin{table*}[!t]
\centering
\captionsetup{justification=centering, labelsep=newline}
\caption{Unmixing performance comparison in terms of SAD for the Cuprite dataset; The best performances are in bold typeface; the second best performances are italicized, and the third best performances are underlined.}
\resizebox{\textwidth}{!}{
\begin{tabular}{c c c c c c c c c}
\hline \hline
Methods 			& \begin{tabular}{c} KbSNMF\\[-0.8ex] fnorm \end{tabular}& \begin{tabular}{c} KbSNMF\\[-0.8ex] div \end{tabular} & $l_{1/2}$-NMF & SGSNMF & \begin{tabular}{c}Min-vol\\[-0.8ex] NMF \end{tabular} & R-CoNMF  & SSRNMF & MVNTF\\
\hline
Alunite 				& 0.4960 & \textit{0.3162} & 0.9145 & 0.7456 & \underline{0.4747} & 0.5745  & \textbf{0.2521} & 	0.3938\\
Andradite 				& \textbf{0.0953} & 0.1977 & 0.5638 & 0.5926 & \textit{0.1161} & 0.2054  & \underline{0.1760} & 0.3118\\
Buddingtonite 				& 0.5837 & 0.2731 & 0.8468 & \textbf{0.1182} & \underline{0.1908} & 0.1936  & \textit{0.1885} & 0.5634\\
Dumortierite 				& 0.4210 & \textit{0.1864} & 0.7002 & 0.7599 & \textbf{0.1316} & \underline{0.2515} &1.0213 & 0.3161	\\
Kaolinite 1 			&	0.4682 & \underline{0.2842} & \textit{0.2366}  & 0.5824 & 0.4536 & 0.5683   &\textbf{0.1375} & 0.4766	\\
Kaolinite 2 				& 0.5530 & \textbf{0.1459} & 0.5869  & 0.7393 & 0.4702 & \underline{0.4079}  &\textit{0.1625} & 0.4595\\
Muscovite 			& \underline{0.2352} & \textbf{0.2086} & 0.4558 & 0.5179 & 0.3358 & 0.3534  & \textit{0.2240} & 0.2555\\
Montmorillonite 			& 0.2926 & 0.4222 & \textbf{0.1127}  & 0.5160 & \underline{0.1651} & 0.2266  & \textit{0.1311} & 0.3130 \\
Nontronite 					& 0.3374 & 0.5450  & 1.0758 & \underline{0.2102} & \textit{0.1845} & 0.4106   & \textbf{0.1249} & 0.2899\\
Pyrope 						& \textit{0.1654} & \underline{0.2630}  & 0.9745 & 0.4753 & 0.4885 & 0.4181  & \textbf{0.0595} & 0.3846\\
Sphene 				& \textbf{0.1590} & \underline{0.3174} & 1.0424 & 0.4875 & \textit{0.2821} & 0.3770  & 0.6160 & 0.4048\\
Chalcedony 				&  \underline{0.2723} & 0.4818 & \textit{0.2291}  & 0.6865 & 0.3511 & 0.5228   & \textbf{0.2394} & 0.3648\\
\hline
Average						& 0.3399 & \textit{0.3035}  & 0.6449 & 0.5360 & \underline{0.3037} & 0.3758  & \textbf{0.2777} &  0.3778\\
\hline \hline
\end{tabular}}
\label{table_performance_cuprite_SAD}
\end{table*}
\subsubsection{Urban dataset} 
\label{section:Urban dataset}
Table \ref{table_performance_urban_SAD} shows SAD values for each of the extracted endmember spectra under the different methods. In terms of SAD, KbSNMF-fnorm outperforms all methods and KbSNMF-div outperforms the rest of the methods except for Min-vol NMF. Also, KbSNMF-fnorm reports the best performance and KbSNMF-div reports the second best performance in extracting the spectra of the endmembers ``Tree'' and ``Roof''. The endmember spectra extracted utilizing KbSNMF-fnorm and KbSNMF-div are shown in Fig. \ref{performance_urban_endmember}, and it can be observed that they closely follow their ground truth spectra. Table \ref{table_performance_urban_RMSE} reports RMSE values for each of the extracted abundance maps, under the different methods. In terms of RMSE, Min-vol NMF outperforms all methods, followed by R-CoNMF and MVNTF. KbSNMF-fnorm reports the best performance in extracting the spectra of the endmembers ``Grass'' and ``Roof', and KbSNMF-div reports the second best performance in extracting the spectra of the endmember ``Asphalt''. The abundance maps extracted utilizing KbSNMF-fnorm and KbSNMF-div are shown in Fig. \ref{performance_urban_abundance}, and it can be observed that they closely follow their ground truth abundance maps.

\subsubsection{Cuprite dataset} 
\label{section:Cuprite dataset}
Table \ref{table_performance_cuprite_SAD} shows SAD values for each of the extracted endmember spectra under the different methods. In terms of SAD, KbSNMF-div sits at the second place while SSRNMF stands at the top. The both KbSNMF forms report compelling results as they show best performance in extracting several endmembers, \textit{i.e} ``Andradite'', ``Kaolinite 2'',  ``Muscovite'', and ''Sphene''. The endmember spectra extracted utilizing KbSNMF-fnorm and KbSNMF-div are shown in Fig. \ref{performance_cuprite_endmember}, and it can be observed that they closely follow their ground truth spectra.

\section{Conclusion}
\label{section:Conclusion}

This paper proposed a blind HU algorithm called KbSNMF, which is based on incorporating the independence of endmembers to the conventional NMF framework. This was done by introducing a novel kurtosis regularizer based on the fourth central moment of a signal which signifies the statistical independence of the underlying signal. We illustrated a comprehensive derivation of the proposed algorithm in this paper along with its performance evaluation in simulated as well as real environments (diverse simulated HSI datasets and three standard real HSI datasets). We have assessed the sensitivity of the proposed algorithm to control parameters, noise levels, number of spectral bands, number of pixels, and number of endmembers of the HSI. We have also provided performance comparisons of the proposed algorithm with the state-of-the-art NMF-based blind HU baselines. Moreover, experimental results verify that dominant performance in endmember extraction can be obtained through the novel algorithm. Hence, the proposed algorithm can be effectively utilized to extract accurate endmembers which can thereafter be passed through as supervisory input data to modern DL methods.

\clearpage

\doublespacing
\bibliographystyle{IEEEtran}
\bibliography{IEEEabrv,mybibfile}

\clearpage

\singlespacing
\begin{IEEEbiography}[{\includegraphics[width=1in,height=1.25in,clip]{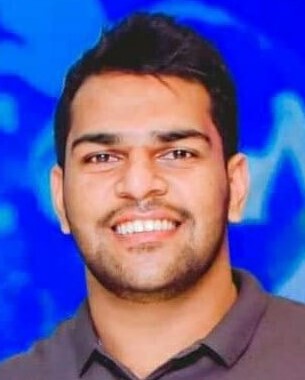}}]{E.M.M.B. Ekanayake }	(Graduate Student Member, IEEE) received the B.Sc. Engineering degree in Electrical and Electronic Engineering from the University of Peradeniya, Sri Lanka, in 2019. His undergraduate research on Hyperspectral Image Processing won the Merit Award for Manamperi Award (Engineering) in 2019 which is presented to the best undergraduate Engineering research project submitted to a Sri Lankan University. He pursued his research on Hyperspectral Image Processing as a Research Assistant at the Office of Research and Innovation Services (ORIS), Sri Lanka Technological Campus, Padukka, Sri Lanka during 2019/2020. He is currently reading for his Ph.D. at the Department of Electrical and Computer Systems Engineering, Monash University, Australia. His previous works have been published in IEEE-TGRS and several other IEEE-GRSS conferences including WHISPERS and IGARSS. He is a Graduate Student Member of the IEEE. His research interests include hyperspectral image processing, remote sensing, machine learning, and computer vision.
\end{IEEEbiography}

\vskip -30pt plus -1fil

\begin{IEEEbiography}[{\includegraphics[width=1in,height=1.25in,clip,keepaspectratio]{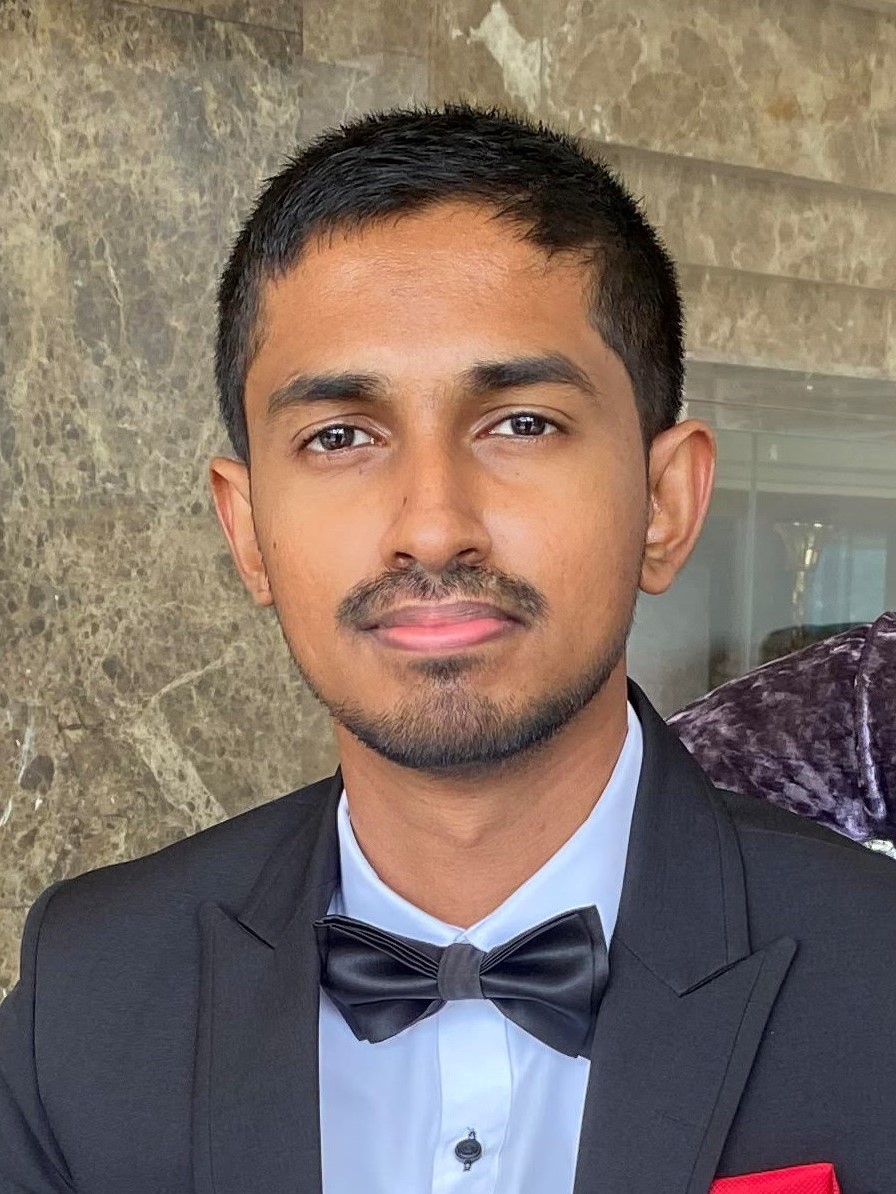}}]{H.M.H.K. Weerasooriya} obtained his degree in Electrical and Electronic Engineering with a first class honours and he currently works as an Instructor in the Department of Electronic and Electrical Engineering. Currently, he is involved in the researches on hyperspectral imaging for remote sensing and agriculture applications, and he has numerous publications in IEEE conferences. His research interests include image processing, signal processing, communication, machine learning and deep learning.
\end{IEEEbiography}

\vskip -30pt plus -1fil

\begin{IEEEbiography}[{\includegraphics[width=1in,height=1.25in,clip]{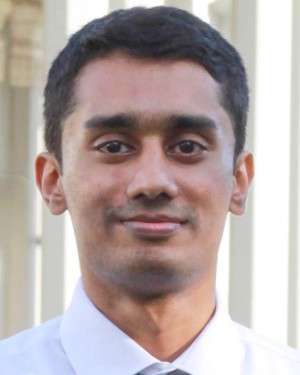}}]{D.Y.L. Ranasinghe} received his B.Sc. Engineering degree in Electrical and Electronic Engineering from the University of Peradeniya, Sri Lanka, in 2020. Immediately after, he joined the School of Engineering, Sri Lanka Technological Campus, Padukka, Sri Lanka as a Research Assistant. His research interests include hyperspectral and multispectral image analysis and processing, blind source separation, and deep learning. He has numerous publications in IEEE conferences.     
\end{IEEEbiography}

\vskip -30pt plus -1fil

\begin{IEEEbiography}[{\includegraphics[width=1in,height=1.25in,clip]{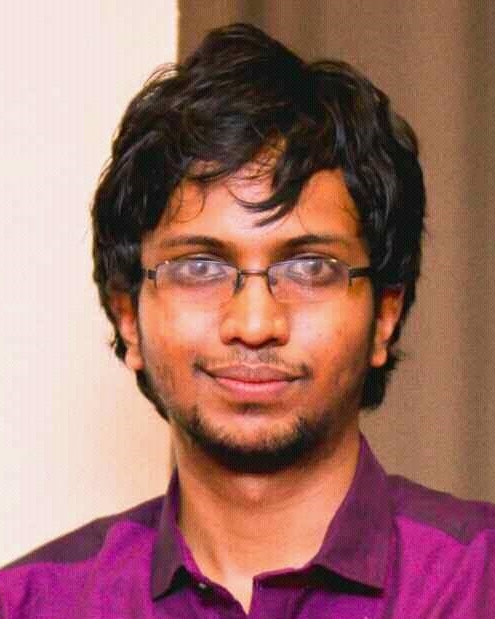}}]{S. Herath} received his B.Sc. Engineering degree in Electrical and Electronic Engineering from the University of Peradeniya, Sri Lanka, in 2020. Immediately after, he joined the Department of Engineering Mathematics, University of Peradeniya as a Teaching Instructor. His research interests include computer vision, image and signal processing, pattern recognition, blind source separation, and machine learning. He has numerous publications in IEEE conferences.      
\end{IEEEbiography}

\vskip -30pt plus -1fil

\begin{IEEEbiography}[{\includegraphics[width=1in,height=1.25in,clip]{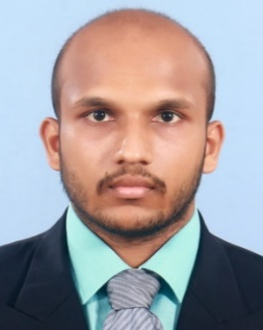}}]{B. Rathnayake} received his B.Sc. Engineering degree in Electrical and Electronic Engineering from the University of Peradeniya, Sri Lanka, in 2017. Immediately after, he joined the Office of Research and Innovation Services (ORIS), Sri Lanka Technological Campus, Padukka, Sri Lanka as a Research Assistant. He is currently working as a Research Assistant at Rensselaer Polytechnic Institute, USA. His research interests include hyperspectral image analysis and processing, graph signal processing, and blind source separation. His previous works have been published in IEEE-TGRS and several other IEEE-GRSS conferences including IGARSS.  
\end{IEEEbiography}

\vskip -30pt plus -1fil

\begin{IEEEbiography}[{\includegraphics[width=1in,height=1.25in,clip]{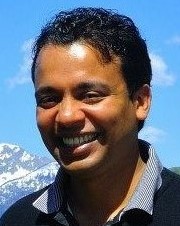}}]{G.M.R.I. Godaliyadda} (Senior Member, IEEE) obtained his B.Sc. Engineering degree in Electrical and Electronic Engineering from the University of Peradeniya, Sri Lanka, in 2005, and Ph.D. from the National University of Singapore in 2011. Currently, he is attached to the University of Peradeniya, Faculty of Engineering, Department of Electrical and Electronic Engineering as a Senior Lecturer. His current research interests include image and signal processing, pattern recognition, computer vision, machine learning, smart grid, bio-medical and remote sensing applications and algorithms. He is a Senior Member of the IEEE. He is a recipient of the Sri Lanka President's Award for Scientific Publications for 2018 and 2019. He is the recipient of multiple grants through the National Science Foundation (NSF) for research activities. His previous works have been published in IEEE-TGRS and several other IEEE-GRSS conferences including WHISPERS and IGARSS. He also has numerous publications in many other IEEE transactions, Elsevier and IET journals and is the recipient of multiple best paper awards from international conferences for his work.
\end{IEEEbiography}

\vskip -30pt plus -1fil

\begin{IEEEbiography}[{\includegraphics[width=1in,height=1.25in,clip]{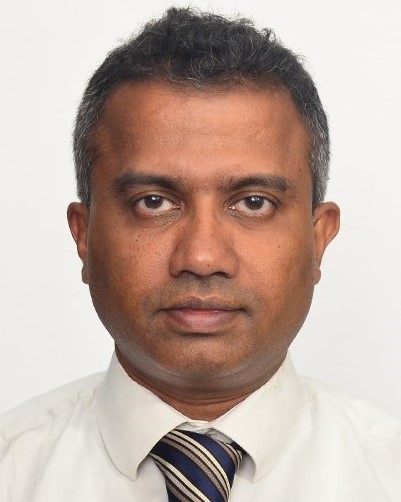}}]{H.M.V.R. Herath} (Senior Member, IEEE) received the B.Sc.Eng. degree in electrical and electronic engineering with 1st class honours from the University of Peradeniya, Peradeniya, Sri Lanka, in 1998, M.Sc. degree in electrical and computer engineering with the award of academic merit from the University of Miami, USA in 2002, and Ph.D. degree in electrical engineering from the University of Paderborn, Germany in 2009. In 2009, he joined the Department of Electrical and Electronic Engineering, University of Peradeniya, as a Senior Lecturer. His current research interests include hyperspectral imaging for remote sensing, multispectral imaging for food quality assessment, Coherent optical communications and integrated electronics. Dr. Herath was a member of one of the teams that for the first time successfully demonstrated coherent optical transmission with QPSK and polarization multiplexing. He is a member of the Institution of Engineers, Sri Lanka and The Optical Society. He is a Senior Member of the IEEE. He was the General Chair of the IEEE International Conference on Industrial and Information Systems (ICIIS) 2013 held in Kandy, Sri Lanka. His previous works have been published in IEEE-TGRS and several other IEEE-GRSS conferences including WHISPERS and IGARSS. He received the paper award in the ICTer 2017 conference held in Colombo Sri Lanka. Dr. Herath is a recipient of Sri Lanka President's Award for scientific research in 2013.
\end{IEEEbiography}

\vskip -30pt plus -1fil

\begin{IEEEbiography}[{\includegraphics[width=1in,height=1.25in,clip]{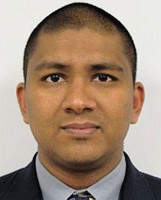}}]{M.P.B. Ekanayake} (Senior Member, IEEE) received his B.Sc. Engineering degree in Electrical and Electronic Engineering from University of Peradeniya, Sri Lanka, in 2006, and Ph.D. from Texas Tech University in 2011. Currently, he is attached to the University of Peradeniya as a Senior Lecturer. His current research interests include applications of signal processing and system modeling in remote sensing, hyperspectral imaging, and smart grid. He is a Senior Member of the IEEE. He is a recipient of the Sri Lanka President's Award for Scientific Publications in 2018 and 2019. He has obtained several grants through the National Science Foundation (NSF) for research projects. His previous works have been published in IEEE-TGRS and several other IEEE-GRSS conferences including WHISPERS and IGARSS. He also has multiple publications in many IEEE transactions, Elsevier and IET journals and has been awarded several best paper awards in international conferences.
	
\end{IEEEbiography}

\clearpage

\end{document}